\documentclass[10pt,pra,aps,twocolumn,showpacs,amsmath,amssymb, floatfix,nofootinbib]{revtex4-1}
\usepackage{amssymb,amsmath,amstext}
\usepackage{url,hyperref,microtype}
\usepackage{graphicx}

\begin{document}

\title[Tensor networks for \texorpdfstring{$p$}{p}-spin models]{Tensor networks for \texorpdfstring{$p$}{p}-spin models} 

\author{Benjamin Lanthier}
\author{Jeremy C\^{o}t\'{e}}
\author{Stefanos Kourtis}

\affiliation{D\'{e}partement de physique \& Institut quantique, Universit\'{e} de Sherbrooke, Sherbrooke, Qu\'{e}bec, J1K 2R1, Canada}

\date{\rm\today}

\begin{abstract}
    We introduce a tensor network algorithm for the solution of $p$-spin models. We show that bond compression through rank-revealing decompositions performed during the tensor network contraction resolves logical redundancies in the system exactly and is thus lossless, yet leads to qualitative changes in runtime scaling in different regimes of the model. First, we find that bond compression emulates the so-called leaf-removal algorithm, solving the problem efficiently in the ``easy'' phase. Past a dynamical phase transition, we observe superpolynomial runtimes, reflecting the appearance of a core component. We then develop a graphical method to study the scaling of contraction for a minimal ensemble of core-only instances. We find subexponential scaling, improving on the exponential scaling that occurs without compression. Our results suggest that our tensor network algorithm subsumes the classical leaf removal algorithm and simplifies redundancies in the $p$-spin model through lossless compression, all without explicit knowledge of the problem's structure.
\end{abstract}

\maketitle

\section{Introduction}
Spin glass physics appears in disciplines far-removed from its origin in condensed matter, including theoretical computer science~\cite{kirkpatrick1985configuration}, biology~\cite{bryngelson1987spin}, and machine learning~\cite{venkataraman1991spin}.
Spin glass models are generally easy to describe, yet hard to solve. One reason is that such models exhibit rugged energy landscapes~\cite{stein2013spin}, trapping optimization algorithms in local minima and leading to exponentially long run times.

A notable counterexample is the $p$-spin model~\cite{mezard_alternative_2002}, which is in fact easy to solve~\cite{ricci-tersenghi_being_2010}. By mapping the model to a linear system of equations modulo 2, Gaussian elimination (GE) allows one to obtain the zero-temperature partition function of the model in polynomial time. While this model is a restricted version of a general spin glass model, its tractable analysis provides useful insights into the physics of spin glasses. Yet the $p$-spin model also exhibits rugged energy landscapes in certain regimes of the parameters, which is why it is a standard benchmark for classical~\cite{bernaschi2021we, kanao2022simulated, aadit2023all} and quantum~\cite{jorg2010first, farhi2012performance, hen2019equation, bellitti2021entropic, kowalsky20223, patil_obstacles_2019} optimization algorithms. In these regimes, simulated annealing fails or is inefficient for any $p>2$~\cite{patil_obstacles_2019, bellitti2021entropic}, and the same is true for quantum annealing~\cite{farhi2012performance}, even when no phase transition is encountered~\cite{patil_obstacles_2019}. Boolean satisfiability and local solvers also struggle with these models~\cite{haanpaa2006hard, jarvisalo2006further,jia2004spin, barthel2002hiding, ricci2001simplest, guidetti2011complexity}.

In this work, we introduce a tensor network algorithm for solving $p$-spin models. A tensor network (TN) is a data structure that allows for compact representation of a given (weighted) graphical model, including (quantum) spin Hamiltonians and constraint satisfaction problems, and whose \emph{contraction} amounts to a (weighted) count of the solutions to the model~\cite{garcia-saez_exact_2011, biamonte2015tensor, kourtis_fast_2019, Meichanetzidis_2019, de_beaudrap_tensor_2021}. While exact TN contraction is computationally hard in general even for restricted graph classes, such as planar grids~\cite{schuch2007computational}, techniques involving tensor \emph{compression} can lead to accurate and efficient approximate estimation of classical partition functions or quantum expectations in specific cases~\cite{evenbly2015tensor, evenbly2017algorithms, gray_hyper-optimized_2022}. TN methods have also previously been used in mean-field studies of graphical models and disordered spin Hamiltonians~\cite{alkabetz2021tensor, pancotti2023one}.

Here, we show that compressed TN contraction applied to the $p$-spin model automatically emulates previously discovered algorithms for the solution of the model in its different phases. In contrast to previous works, the compression we perform is \emph{exact}, meaning that it only resolves and simplifies redundancies in the TN at each step without loss of information. We illustrate the above with an application to the $3$-spin model, in which the average number of interactions per spin $\alpha$ controls transitions to different thermodynamic phases in the structure of the problem~\cite{mezard_alternative_2002}. We find that compressed TN contraction automatically implements the leaf removal algorithm~\cite{mezard_alternative_2002} and thus efficiently solves the problem when $\alpha < \alpha_d$, at which point a dynamical transition occurs. In contrast, compressed TN contraction scales superpolynomially when $\alpha > \alpha_d$ but improves substantially on the exponential scaling of TN contraction without compression. We further show that when $\alpha \in \left[2/3, 3/4 \right]$, compressed TN contraction outperforms naive GE. Finally, by devising a graphical scheme that exactly captures the dynamics of compressed TN contraction in the special case of spins appearing in exactly \emph{two} interaction terms, for which no leaf removal occurs, we show numerically that the TN method solves the problem in \emph{subexponential} time.

\section{Definitions}
\subsection{The \texorpdfstring{$p$}{p}-spin model} \label{sec:p-spin}
We can write the $p$-spin model by specifying a bipartite graph $G = (U,V,E)$, where $U$ is the set of nodes representing the $n = |U|$ spins, $V$ is the set of nodes representing the $m = |V|$ interaction terms, and $E$ is the set of edges connecting spin nodes to interaction nodes.
We can then write the Hamiltonian of the $p$-spin model as:
\begin{equation} \label{eq:hamiltonian}
    H = \frac{1}{2} \left[m -\sum_{v \in V} J_{v}\prod_{u \in N(v)} \sigma_{u} \right],
\end{equation}
where $J_v \in \left\{-1, 1\right\}$ are the couplings for the interaction at node $v$, $\sigma_u \in \left\{-1,1\right\}$ is the value of the spin at node $u$, and $N(v)$ is the set of $p$ neighbours for the interaction described by node $v$.
The minimum energy is zero, and it occurs when every interaction satisfies $J_{v} = \prod_{u \in N(v)} \sigma_{u}$.
In this paper, we are interested in counting the number of zero-energy configurations for a given ensemble of bipartite graphs, that is, evaluating the zero-temperature partition function of the model.

By letting $\sigma_u = (-1)^{x_u}$ and $J_v = (-1)^{b_v}$, we can rewrite the search for zero-energy configurations from Eq.~\ref{eq:hamiltonian} as
\begin{equation} \label{eq:Ax=b}
    A\vec{x} = \vec{b} \mod{2},
\end{equation}
where $A \in \{0, 1\}^{m \times n}$ is the biadjacency matrix of the graph $G$, with $A_{vu} = 1$ indicating $u \in N(v)$ and zero otherwise, $\vec{x} \in \left\{ 0,1 \right\}^n$ encodes the spin configuration, and $\vec{b} \in \left\{ 0,1 \right\}^m$ encodes the couplings. 
Finding the zero-energy configurations for Eq.~\ref{eq:hamiltonian} is equivalent to solving the matrix equation~\ref{eq:Ax=b}. 
Counting the number of configurations also involves manipulating Eq.~\ref{eq:Ax=b}. With this form, we can then cast the problem into the the language of Boolean satisfiability (SAT), which we detail below.

\subsection{The \#\texorpdfstring{$p$}{p}-XORSAT problem}\label{pxorsat_problem}

\subsubsection{Definition}\label{xorsat_formal_definition}
In its most general form, a SAT problem is the problem of deciding whether a logic formula built from a set of boolean variables $\{x\} = \{x_1, x_2, ..., x_n\}$ and the operators $\wedge$ (conjunction), $\vee$ (disjunction), and $\neg$ (negation) evaluates to true, i.e., is \textit{satisfiable}~\cite{garey1979computers}.
The SAT problem is characterized by the conjunction of clauses, each comprising disjunctions of variables where the negation operator may be applied.
The SAT problem is NP-complete, and the same is true for many of its variants.

The constraint stipulating that every clause must consist of exactly $p$ variables defines the $p$-SAT problem, which is also NP-complete.
Counting the number of solutions that satisfy a given SAT problem, if any exist, defines the \#SAT problem, which is even more challenging, falling under the \#P-complete class.
This property extends to \#$p$-SAT problems for $p \geq 2$.

The variant of the \#$p$-SAT problem that lets us count the number of zero-energy configurations of a given $p$-spin model is the \#$p$-XORSAT problem, defined below.

The \#$p$-XORSAT problem requires only a modification of the operators within the clauses from the standard $p$-SAT formulation.
The disjunction is replaced by the \textit{exclusive-or} (XOR) operator, which is mathematically represented by the summation modulo $2$ operator ($\oplus$). Given $A$ and $\vec{b}$ as in Eq.~\ref{eq:Ax=b}, we can define an instance $\phi$ of the $p$-XORSAT problem as:
\begin{equation}\label{eq:xorsat_def}
    \begin{split}
        \phi(\{x\}) &= \bigwedge_{i = 1}^m c_i,\\
        c_i &= 1 \oplus b_i \oplus A_i \cdot \vec{x},\\
        \vec{x} &= \left(x_1, x_2, \ldots, x_n \right) \in \{0,1\}^n,
    \end{split}
\end{equation}
where $A_i \in \left\{ 0,1 \right\}^n$ is the $i$-th row of $A$ and $b_i$ is the $i$-th component of $\vec{b}$, $A_i \cdot \vec{x}$ indicates the dot product between $A_i$ and $\vec{x}$ (modulo $2$), and $c_i = 1$ implies the clause is satisfied ($b_i \oplus A_i \cdot \vec{x} = 0$).

When one generates $A$ by placing $p$ ones in each row uniformly at random with no repeated rows and uniformly chooses $\vec{b} \in \left\{0,1 \right\}^m$, the clause density $\alpha \equiv m / n$ characterizes much of the problem. In particular, $p$-XORSAT has two phase transitions~\cite{mezard_alternative_2002}. The first occurs at $\alpha_d$, which indicates a dynamical transition in the structure of the solution space by dividing solutions into well-separated (in Hamming distance) clusters. The second occurs at the critical transition $\alpha_c$, where, with high probability, any instance becomes unsatisfiable (no solutions). This point signifies a similar phase transition even when $\vec{b} = 0$, meaning the configuration $\vec{x} = 0$ is always a solution~\cite{ricci2001simplest}. For $p = 3$, the constants are $\alpha_d \approx 0.818$ and $\alpha_c \approx 0.918$~\cite{mezard_alternative_2002}.

\subsubsection{Gaussian elimination}
Given a $p$-XORSAT instance $\phi \left( \left\{x\right\} \right)$, we first translate it into the form of Eq.~\ref{eq:Ax=b}.
Then, we apply GE on the augmented matrix $[A|\vec{b}]$.
If the system is inconsistent, there are no solutions. Otherwise, the solution count is:
\begin{equation}\label{eq:ge}
    \text{\#Solutions} = 2^{n - \mathrm{rank}(A)},
\end{equation}
where all operations are modulo $2$, as in applying GE. \#$p$-XORSAT is thus in P since it can be solved in at most $\mathcal{O}(n^3)$ time and $\mathcal{O}(n^2)$ memory.

In Ref.~\cite{braunstein_complexity_2002}, the authors studied the time and memory requirements for solving Eq.~\ref{eq:Ax=b} for $p = 3$ using a ``simple'' version of GE. This version solves the linear equations in the order they appear in Eq.~\ref{eq:Ax=b} and with respect to a random variable. The authors showed that this simple algorithm will solve the problem in $\propto n$ time and memory when $\alpha \leq 2/3$, and in $\propto n^3$ time and $\propto n^2$ memory when $\alpha > 2/3$.

The authors also presented a ``smart'' version
of GE, where one first looks for the variable appearing in the least number of equations left to be solved (ties broken arbitrarily), then solves for that variable and substitutes it into the remaining equations. They argued that this smarter version of GE will solve the problem in $\propto n$ time and memory when $\alpha < \alpha_d$, and in $\propto n^3$ time and $\propto n^2$ memory when $\alpha > \alpha_d$.

When one solves an equation that contains a variable which only appears in that equation, one can interpret the process graphically as a ``leaf removal'' algorithm~\cite{mezard_alternative_2002}. We describe it below because it provides intuition as to why the ``smart'' version of GE is more efficient and will help explain the behaviour of our TN algorithm.

\subsubsection{Leaf removal}\label{sec:leaf-removal-algorithm}
Suppose we have an instance for $p=3$ and the variable $x$ only appears in the linear equation $x \oplus y \oplus z = b$. No matter what values $y$ and $z$ take, it is always possible to choose $x$ to make the equation true. We can therefore solve this equation for $x$, and only fix it once we have solved the rest of the (fewer) linear equations. But removing this equation may now cause $y$ or $z$ to only appear in a single other equation, so we solve those equations for $y$ and $z$, and then what remains is an even smaller linear system. The process will continue until the remaining variables participate in at least two equations. In terms of the matrix $A$ in Eq.~\ref{eq:Ax=b}, each column will have at least two 1s. (Note that if a variable appears in no equations it is, in essence, not part of the problem and so we can ignore it and simply multiply the count by 2.)

This algorithm is called leaf removal~\cite{mezard_alternative_2002}, and it allows us to simplify the $p$-XORSAT problem. Graphically, the algorithm begins with the bipartite graph $G$ representing the problem, then iteratively finds variable nodes $u \in U$ such that $\mathrm{deg}(u) = 1$, and deletes the clause node $v \in N(u)$ and $v$'s associated edges. The algorithm continues until either no clause nodes remain (and therefore, no edges) or a ``core'' remains, a subgraph of $G$ where each variable node has degree at least two. One can then construct a solution to the original formula by working backwards from a solution to the formula corresponding to the core graph.

In Ref.~\cite{mezard_alternative_2002}, the authors showed that, for the ensemble where $p = 3$ and one picks each clause uniformly at random from the $\binom{n}{3}$ distinct tuples of variables, the leaf algorithm will succeed in reducing the corresponding graph to the empty graph when $\alpha < \alpha_d \approx 0.818$. Because at each step of the algorithm one can fix a variable node of degree $1$ in order to remove a clause node, when no core remains the count will be $2^{n-m}$, where $m$ is the number of clauses (or variables we have fixed). When $\alpha > \alpha_d$, a core will remain, which means leaf removal is not enough to solve the entire problem. The value $\alpha_d$ indicates a dynamical transition in the problem, and it corresponds to a change in the structure of the set of solutions. The ``smart'' GE uses this principle to achieve a speedup over the standard version.

We also note that when no core remains at the end of leaf removal, one can interpret the algorithm as finding a permutation of the rows and columns of the matrix $A$ such that one can transform $A$ into triangular form. Suppose the variable $x_i$ only appears in equation $j$. One would then permute the rows $1$ and $j$ of $A$, as well as the columns $1$ and $i$. Ignoring the first row and column of $A$, repeat the same procedure. Continuing in this way will yield a matrix $A'$ which is in triangular form and has the same rank as $A$. The triangular form of $A'$ implies that its rank is simply the number of rows, allowing one to calculate the number of solutions.

In the case of the $p$-XORSAT problem, this algorithm demonstrates that we can graphically identify and eliminate redundancy, reducing the problem's size by focusing on the remaining core. Graphically, this problem does not only exhibit this rank-$1$ variable redundancy; two more are explained in the following section.

\subsubsection{Graphical simplifications} \label{sec:XORSAT-simplifications}
There exist graphical rules, such as the leaf removal explained in Sec.~\ref{sec:leaf-removal-algorithm}, that let us simplify a $p$-XORSAT problem. These will be used in Sec.~\ref{sec:graphical-method}, where we develop a complementary graphical method for TN contraction. Note that we will study the case where $\vec{b} = 0$ for simplicity.
Then, we have the following examples of simplifications.

\begin{figure}[htbp]
    \centering
    \includegraphics[width=0.49\textwidth]{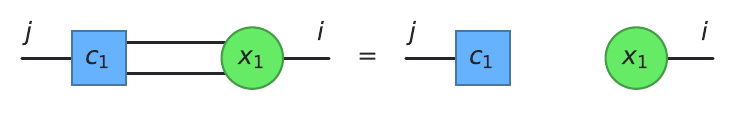}
    \caption{Graphical representation of the Hopf law. Clause nodes are blue squares, and variable nodes are green circles.}
    \label{fig:hopf_law}
\end{figure}

The first example is the Hopf law~\cite{denny_algebraically_2012}, where a clause involves the same variable multiple times. In this case, since $i \oplus i = 0$ for boolean indices, when there are $t$ occurrences of a variable in a clause, only $t \mod{2}$ of them are necessary and the rest are redundant. In Fig.~\ref{fig:hopf_law}, we show an example for $t = 2$.

\begin{figure}[htbp]
    \centering
    \includegraphics[width=0.49\textwidth]{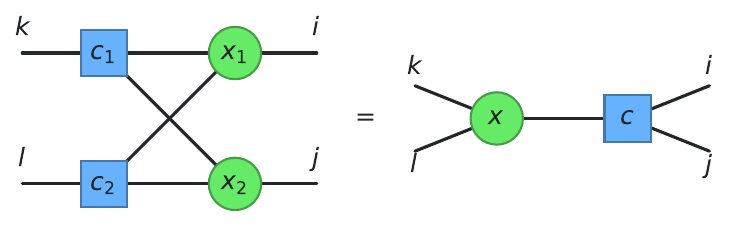}
    \caption{Graphical representation of the bialgebra law. Clause nodes are blue squares, and variable nodes are green circles.}
    \label{fig:bialgebra_law}
\end{figure}

The second example is the bialgebra law~\cite{denny_algebraically_2012}, where a set of clause nodes are all connected to a set of variable nodes. An example for two clauses and two variables is shown in Fig.~\ref{fig:bialgebra_law}. These structures simplify to a single clause and single variable, as shown in Fig.~\ref{fig:bialgebra_law}.

These simplifications correspond to eliminating redundancies in the problem. Resolving these redundancies can be exploited to solve the problem faster.

\subsection{Tensor networks}

TNs are a data structure that encodes a list of tensor multiplications. Intuitively, one can imagine a TN as a graph where each node represents a tensor, and edges represent the common axes along which one multiplies two tensors~\footnote{Though this does not factor into our work, it is also possible to have ``free'' edges with only one end connected to a node, indicating an axis in which no tensor multiplication occurs.}. By contracting together neighboring nodes---multiplying the corresponding tensors together---one can sometimes efficiently compute a variety of quantities, making it a useful numerical method. Originally developed to efficiently evaluate quantum expectation values and partition functions of many-body systems, this tool now has applicability in many domains, including quantum circuit simulation~\cite{seitz_simulating_2023} and machine learning~\cite{wang_tensor_2023}. As shown in~\cite{garcia-saez_exact_2011}, this tool can also be used for $p$-SAT problems.

For our work, contracting all of the tensors in the network together will yield the number of solutions to Equation~\ref{eq:Ax=b}. Below, we review the main ideas for TN methods that are relevant for us and determine the performance of our algorithm. These elements are: how to perform contractions, the importance of contraction ordering, and how to locally optimize the sizes of the tensors (which affect the memory requirements). We then describe our TN algorithm for the \#$p$-XORSAT problem in Sec.~\ref{sec:tn-for-XORSAT}.

\subsubsection{Contraction} \label{sec:contraction}

A single tensor is a multidimensional array of values. Graphically, the number of axes (or rank) of the tensor is the degree of the corresponding node, and the size of the tensor is the number of elements (the product of the dimensions of the axes). The size of the TN is then the sum of all the tensor sizes. For any TN algorithm, one must keep track of the size of the TN to ensure the memory requirements do not exceed one's computational limits. In particular, one must consider how contracting tensors together changes the TN's size.

\begin{figure}[htbp]
    \begin{center}
    \includegraphics[width=0.5\textwidth]{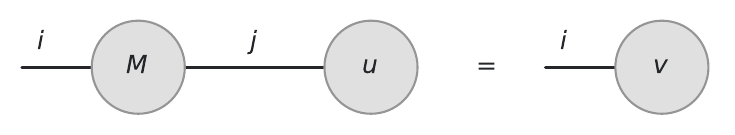}
    \end{center}
    \caption{Matrix-vector multiplication in TN format.}
    \label{fig:tn_mat-vec_multiplication}
\end{figure}

A simple example of contraction is the matrix-vector multiplication, which is represented graphically in Fig.~\ref{fig:tn_mat-vec_multiplication}.
Here, the vector $\vec{u}$ (a rank-$1$ tensor) is represented by a node with a single edge connected to it and the matrix $M$ (a rank-$2$ tensor) is also a node, but with two edges. The matrix-vector multiplication shown in Fig.~\ref{fig:tn_mat-vec_multiplication} can also be written as the following summation:
\begin{equation}\label{eq:einsum}
    \sum_j M_{ij}u_j = v_i.
\end{equation}
In general, one can write the contraction of a TN by this summation over all the common (shared) axes. We will sometimes call tensors with common axes \emph{adjacent}, in reference to a TN's graphical depiction.

When contracting tensors where each axis has the same dimension, we can graphically determine the resulting size by looking at the degree of the new node. In Fig.\ref{fig:tn_mat-vec_multiplication}, the resulting tensor has rank 1, which is the same as $\vec{u}$'s rank. However, the resulting tensor size can be much larger than the original tensors. Suppose we contract tensors of rank $d_1$ and $d_2$ which share a single common axis and each axis has dimension $2$, then the size of the resulting tensor will be $2^{d_1 + d_2 - 2}$ and thus scales exponentially in tensor ranks.

\subsubsection{Contraction order}\label{sec:contraction-ordering}
Though we can carry out the contraction of a TN in any order, the size of the TN in intermediate steps of the contraction can vary widely. Ideally, a contraction will choose an order that limits the memory required to store the TN during all steps of the contraction, making it feasible. Given a contraction order, we can define the \emph{contraction width} $W$ of the TN~\cite{gray_hyper-optimized_2021} in two equivalent ways:
\begin{equation} \label{eq:contraction-width}
    W =
    \begin{cases}
        \max_{v \in P} \mathrm{deg}(v) & \text{(graphical)},\\
        \max_{T \in \mathcal{T}} \log_2{s(T)} & \text{(tensors)}.\\
    \end{cases}
\end{equation}
For the graphical representation, $P$ is the set of nodes representing the tensors present at any stage of the contraction. In the tensor representation, $\mathcal{T}$ is the set of all tensors that are present at any stage of the contraction, and $s(T)$ is the size (number of elements) of the tensor $T$. Note that $W$ depends on the TN and the contraction order. Then, up to a prefactor~\cite{gray_hyper-optimized_2021}, $2^W$ captures the memory requirements for the entire contraction. We use the contraction width as a proxy to runtime because it defines the largest tensor that one must manipulate during the contraction using multilinear operations, which take polynomial time in the size of that tensor~\cite{gray_hyper-optimized_2021}.
Finding such orderings is an optimization problem and algorithms exist to find optimized contraction ordering according to the TN structure.
While finding the optimal contraction order is easy in some cases, for example, a square lattice, it is much more complex in others, such as random networks~\cite{gray_hyper-optimized_2022}.
In general, the computational demands of a TN contraction grow exponentially with the number of tensors in both time and memory.
Even so, a method called bond compression allows us to further optimize the contraction by accepting a little error. We review this method below, and we explain in Sec.~\ref{sec:XORSAT-simplifications} how we use bond compression in a novel way.

\subsubsection{Bond compression}\label{Compression}
Bond compression involves, in its simplest form, performing a contraction-decomposition operation on adjacent tensors within the TN. The term ``bond'' refers to the common index between tensors.
The decomposition step primarily uses rank-revealing methods such as QR or \textit{singular value decomposition} (SVD).
Of these, the SVD plays a central role in TN algorithms.
By setting a threshold value for singular values, either absolute or relative, we retain only the singular values above the threshold and corresponding singular vectors, thereby approximating subsequent contractions.
This approach facilitates the contraction of larger TNs by reducing the contraction width during the process.
However, in general, this comes at the expense of approximating the final result.

We implement bond compression as follows.
Given two adjacent tensors $T_A$ and $T_B$ in the network, we transform them into the approximate tensors $\tilde{T}_A$ and $\tilde{T}_B$ as
\begin{equation}\label{eq:compression_schedule}
    \begin{split}
        T_AT_B &= Q_AR_AR_BQ_B\\
        &= Q_AR_{AB}Q_B\\
        &= Q_A(U\Sigma V^\dagger)Q_B\\
        &\approx Q_A(\tilde{U}\tilde{\Sigma} \tilde{V}^\dagger)Q_B\\
        &= (Q_A\tilde{U}\tilde{\Sigma}^\frac{1}{2})(\tilde{\Sigma}^\frac{1}{2}\tilde{V}^\dagger Q_B)\\
        &= \tilde{T}_A\tilde{T}_B.
    \end{split}
\end{equation}
The first equality comes after applying a QR decomposition to the tensors.
Since the QR decomposition operates solely on matrices, we first need to reshape those tensors into matrices before decomposing them.
Concretely, if we have a tensor $T$ that has indices $(i_1, i_2, ..., i_k)$ and we want to apply the QR on the index $i_3$, then the reshaping would give a matrix with indices $(\prod_{j \neq 3}i_j, i_3)$ (where the product signifies grouping the indices into a composite index).
This matrix allows for the direct application of the QR decomposition on the desired dimension.
The second equality comes from multiplying the matrices $R_A$ and $R_B$ to get the matrix $R_{AB}$.
The third equality comes after performing the SVD on $R_{AB}$.
Then, the threshold is applied, reducing the sizes of the singular values matrix, of $U$ and of $V$ and possibly approximating the result.
The following equality comes from splitting this diagonal singular values matrix into two equal ones.
The final equality comes from multiplying the matrices together in each parenthesis to get two new tensors with a ``compressed'' bond between them.
This schedule optimizes the bond compression since the contraction between two tensors of possibly high dimensions is avoided.

\section{Methodology}
\subsection{Tensor networks for \texorpdfstring{$p$}{p}-XORSAT}
\label{sec:tn-for-XORSAT}
\begin{figure}[htbp]
    \begin{center}
    \includegraphics[width=0.5\textwidth]{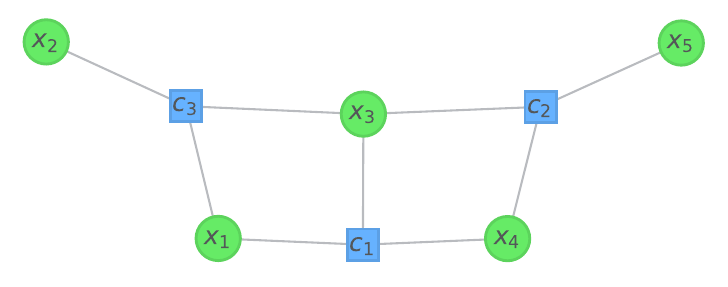}
    \end{center}
    \caption{An example of a TN representing a $3$-XORSAT instance with $n = 5$ (green circles), $m = 3$ (blue squares).}
    \label{fig:tn-example}
\end{figure}

As shown in Ref.~\cite{garcia-saez_exact_2011}, we can map any $p$-XORSAT instance as a TN. Contracting it will yield the number of solutions to the problem. As with the $p$-spin model in Sec.~\ref{sec:p-spin}, we can define a $p$-XORSAT instance by a bipartite graph $G = (U,V,E)$ and a vector $\vec{b}$ of parities. Then, to each node $u \in U$ we will assign a ``variable'' (or COPY) tensor, which has the form:
\begin{equation}\label{eq:COPY}
    T^{\text{COPY}\{u\}}_{i_1i_2...i_d} = \begin{cases}
        1, & \text{ if } i_1 = i_2 = ... = i_d,\\
        0, & \text{ else,}\\
    \end{cases}
\end{equation}
where the indices $i_1i_2...i_d$ are boolean and $d = \mathrm{deg}(u)$. For each node $v \in V$, we will assign a ``clause'' (or XOR) tensor of the form:
\begin{equation} \label{eq:XOR-tensor}
    T^{\text{XOR}\{v\}}_{i_1i_2...i_p} = \begin{cases}
        1, & \text{ if } i_1 \oplus i_2 \oplus ... \oplus i_p = b_v\\
        0, & \text{ else }\\
    \end{cases},
\end{equation}
where the indices are also boolean, $p = \mathrm{deg}(v)$ and $b_v$ is the parity associated to clause $v$. Finally, the edges $E$ indicate which indices are common between different tensors in the TN and need to be summed over. Obtaining the solution count for the problem involves writing a summation over all of the common indices, yielding an expression similar (but much more involved for larger TNs) to Eq.~\ref{eq:einsum}.
In Fig.~\ref{fig:tn-example}, we give an example of a $3$-XORSAT instance with $n = |U| = 5$ and $m = |V| = 3$.

As explained in Sec.~\ref{sec:contraction-ordering}, we can evaluate the contraction width $W$ of those TNs by extracting the highest tensor rank reached during its contraction.

\subsection{Eliminating redundancies through bond compression}
There are several possible simplifications for a $p$-XORSAT problem that occur during the intermediate steps of the TN contraction. By recognizing these simplifications, we can reduce the size of the TN and therefore the time and memory requirements for its contraction.
We will focus on the case where $\vec{b} = \vec{0}$, so all parities are even.

We will use bond compression to contract and decompose all adjacent tensors in the TN, a process commonly called a \emph{sweep}, which is standard practice in TN methods. However, we will \emph{not} remove any nonzero singular values in the decomposition. If the tensors are full-rank, this is useless; the tensors remain unchanged after performing bond compression. On the other hand, TNs representing $p$-XORSAT problems often contain redundancy (see Sec.~\ref{sec:XORSAT-simplifications}), which results in singular values that are zero to numerical accuracy. Therefore, performing bond compression and removing null singular values allows us to reduce the tensor sizes while keeping the resulting contraction exact.

\begin{figure}[htbp]
    \begin{center}
    \includegraphics[width=0.5\textwidth]{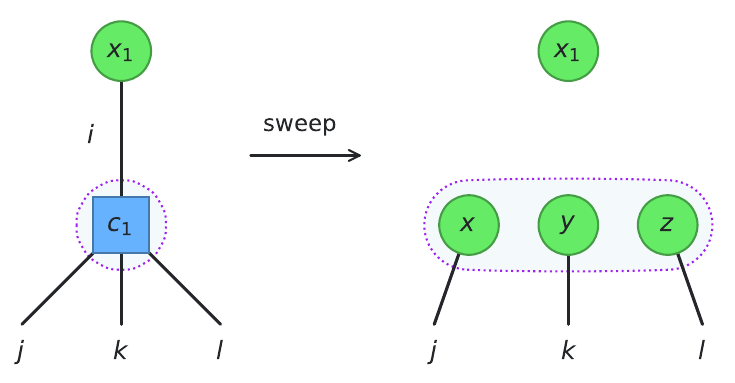}
    \end{center}
    \caption{Applying bond compression on a rank-$1$ variable tensor (green circular node labelled $x_1$ on the left) connected to a rank-$4$ clause tensor (blue square node labelled $c_1$ on the left). The result is a scalar and a rank-$3$ tensor that is equivalent to the tensor product of three rank-$1$ variable tensors.}
    \label{fig:degree1_sweep}
\end{figure}

An interesting fact with this method is that applying bond compression to the bond between a rank-$1$ variable tensor and a rank-$d$ clause tensor will effectively remove the bond, leading to a scalar (rank-$0$ tensor) and a rank-$(d-1)$ tensor.
This rank-$(d-1)$ tensor will be composed of only ones (with a prefactor), which is equivalent to the tensor product of $d-1$ rank-$1$ variable tensors. We illustrate an example of this in Fig.~\ref{fig:degree1_sweep}.
The following sweep step will then remove those $d-1$ bonds (because they connect to a rank-$1$ variable, or COPY, tensor). This means the algorithm effectively removes the clause tensor and all its bonds, which is equivalent to one step in the leaf removal algorithm. This process could cascade through the entire TN, potentially eliminating all its bonds or resulting in a leafless core, giving the same outcome as the leaf removal algorithm. Therefore, bond compression sweeps automatically implement the leaf removal algorithm.

The contraction width will be the figure of merit for the performance of this algorithm because of its relation with the maximum intermediate tensor size (see Eq.~\ref{eq:contraction-width}).

\subsection{Graphical contraction} \label{sec:graphical-method}
When $\alpha < \alpha_d$, leaf removal is likely to completely simplify the graph encoding the problem (Sec.~\ref{sec:leaf-removal-algorithm}). Translated to TN contraction, the bond compression shown in Fig.~\ref{fig:degree1_sweep} would be enough to dramatically simplify the TN contraction. This allows us to scale our simulations to large system sizes. However, when $\alpha > \alpha_d$, a core will likely remain. In this case, the remaining TN to contract comprises a core, and this will change the scaling of resources. In particular, the presence of a core will increase the contraction width (and therefore the memory requirements) much more quickly than when $\alpha < \alpha_d$. This limits our ability to test the performance of our algorithm on large instances in this regime.

To bypass this bottleneck and provide further scaling evidence, we develop a graphical algorithm that allows us to study the contraction width throughout a contraction by only studying the connectivity of the instance's graph. As discussed in Sec.~\ref{sec:contraction}, this is always possible for any exact contraction of a TN, since one simply needs to keep track of the tensor ranks at each step of the contraction (regardless of the tensors' contents). However, because we seek to study the performance of our TN algorithm that detects simplifications through bond compression, we must also encode the graphical patterns that will lead to simplifications. We will make use of the graphical simplifications discussed in Sec.~\ref{sec:XORSAT-simplifications}, as well as more discussed in Sec.~3 of Ref.~\cite{denny_algebraically_2012}.

\begin{figure}[htbp]
    \centering
    \includegraphics[width=0.5\textwidth]{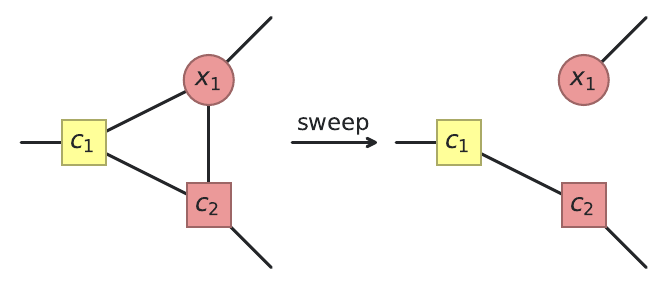}
    \caption{One of the two possible cases of the triangle simplification. Node $c_1$ is in one cluster (yellow), and nodes $(c_2, x_1)$ are in the other (red). There are initially two shared edges between the clusters. After the sweep, edges $c_1x_1$ and $c_2x_1$ disappear, resulting in only one shared edge remaining between the two clusters.}
    \label{fig:triangle_rule}
\end{figure}

The graphical algorithm works as follows. Starting from a graph $G$ encoding the instance, each node will always represent either a variable or a clause, and by default we will assign each node to a distinct ``cluster''. The algorithm ``contracts'' two nodes by assigning them to the same cluster. One can think of the cluster as a contracted tensor. Then, whenever the algorithm performs a ``sweep'', it will search for any possible simplifications \emph{between} clusters involving variable and clause nodes. If the algorithm finds any, it will perform the simplifications by removing edges in the problem~\footnote{The algorithm can also remove edges within a cluster, if it is part of the simplification (see Fig.~\ref{fig:triangle_rule}).}. The algorithm alternates between sweeping and contracting until every node in the graph belongs to the same cluster, in which case it terminates. It uses the same contraction ordering as in our TN algorithm.
In graphical contraction, the goal is to obtain the sizes of intermediate tensors encountered in the contraction, not the values of the tensors themselves. Therefore, the graphical algorithm will not produce a solution count, just a contraction width. We also note that a degree-$2$ variable tensor is, in its tensor representation, equal to a $2 \times 2$ identity matrix (see Eq.~\ref{eq:COPY}). Knowing that, we can replace any degree-$2$ variable nodes in a cluster by edges.

The rank of an intermediate tensor is the number of outgoing edges from a cluster, and its size is:
\begin{equation} \label{eq:size-cluster}
    \text{size}_{\text{cluster}} = 2^{\text{\#outgoing edges}}.
\end{equation}
Taking the maximum number of outgoing edges over all contraction steps and clusters directly yields the contraction width.

We now interpret the sweeping method as implementing graphical simplifications. Recall that the TN contraction is a sum over all the boolean indices of the tensors and only the indices which satisfy the logic of the TN will contribute $1$ to the sum (and $0$ otherwise), yielding the solution count to the problem. Therefore, any simplifications from bond compression must correspond to redundancy in specifying the logic of the TN. Suppose the algorithm is compressing the bonds between tensors $T_A$ and $T_B$. For concreteness, suppose there are $k$ bonds. The algorithm will first transform the $k$ bonds of dimension $2$ into a single bond of dimension $2^k$. Then, the algorithm will compress that bond according to Eq~\ref{eq:compression_schedule}, yielding new tensors $\tilde{T}_A$ and $\tilde{T}_B$ such that their shared bond is minimized due to the SVD. We observe that the new shared bond has dimension $2^{k'}$ for $k' \leq k$, and $k'$ corresponds to the minimum number of bits needed to preserve the logic of contracting $T_A$ and $T_B$. Note that we can interpret a single bond of dimension $2^{k'}$ as $k'$ bonds of dimension $2$, which is how we display our graphical simplifications.

For example, in the leaf removal algorithm, compressing the bond of a rank-$1$ variable tensor $T^{\mathrm{COPY}}$ with a rank-$4$ clause tensor $T^{\mathrm{XOR}}$ will yield a shared ``bond'' of dimension $2^0$, due to redundancy in the representation of contracting those two tensors.
This dimension $1$ ``bond'' signifies that the contraction of those tensors will be a tensor product that reduces to an element wise multiplication of tensor $\tilde{T}^{\mathrm{XOR}}$ with the scalar value of $\tilde{T}^{\mathrm{COPY}}$.
Similarly, we show below that there are several known logical simplifications present between tensors in these TNs which minimize the number of bits needed to preserve the contraction, implying the QR/SVD will find them. We observe as much in our experiments, which led us to developing our graphical algorithm.

The algorithm must detect and simplify any tensor that our TN algorithm would simplify. For the (2,3)-biregular graph ensemble ($\alpha = 2/3$ leaf-free instances) we consider, only a subset of the possible $p$-XORSAT simplifications are present. Following the examples in Ref.~\cite{denny_algebraically_2012}, our graphical algorithm detects the following possible simplifications (we assume $\vec{b} = \vec{0}$ for simplicity):
\begin{itemize}
    \item Fusion rule,
    \item Generalized Hopf law,
    \item Triangle simplification,
    \item Multiple edges between nodes of the same type,
    \item Scalar decomposition.
\end{itemize}

\begin{figure}[htbp]
    \centering
    \includegraphics[width=0.4\textwidth]{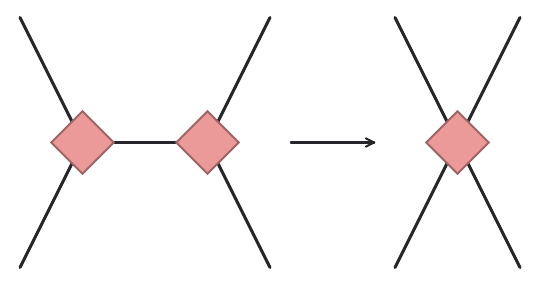}
    \caption{The fusion rule on two nodes that are in the same cluster, identified as red here. Nodes of diamond shape represent nodes that could be either of type clause or of type variable.}
    \label{fig:fusion_rule}
\end{figure}

The \textit{fusion rule} says that neighboring clause nodes in the same cluster can be contracted together to form a bigger clause node, and the same is true for variable nodes. In this case, we actually replace the two nodes with a single node representing them. Their corresponding tensor representations would then be exactly those of a clause or variable tensor of larger rank. This rule is schematically shown in Fig.~\ref{fig:fusion_rule}. One can also apply the same rule for nodes of the same type which share multiple edges. However, for clause nodes, there will be an overall numerical factor of $2^{\#\text{shared edges} - 1}$ in the entries of the tensor, corresponding to the summation over shared indices. Since we are only concerned with the size of the tensors, this coefficient is not relevant.

The \textit{generalized Hopf law} ensures that if a clause node and a variable node share $t$ edges and the degree of each is greater than $t$, a sweep will leave $t\mod{2}$ edges between them (as discussed in Sec.~\ref{sec:XORSAT-simplifications}).

The \textit{triangle simplification} is an implementation of the Hopf law between two clusters that, between them, contain a ``triangle'' of nodes. Those triangles contain two nodes of one type (clause or variable) and one of the other. Because we always contract nodes of the same type within a cluster using the fusion rule, a triangle simplification can only occur when the nodes of the same type are in different clusters. When we sweep between these clusters, applying the fusion rule and then a basic Hopf law will remove edges, as shown in Fig.~\ref{fig:triangle_rule}.

\begin{figure}[htbp]
    \centering
    \includegraphics[width=0.5\textwidth]{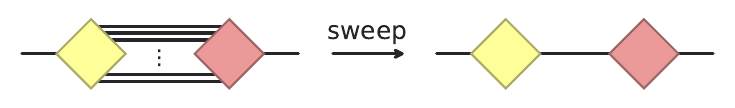}
    \caption{The multiple edges between nodes of the same type simplification. The nodes are in different clusters (yellow and red), and initially share multiple edges. After a sweep, only one edge is needed to represent the same tensor structure.}
    \label{fig:same_type_simplification}
\end{figure}

The simplification of \textit{multiple edges between nodes of the same type} is a variant of the fusion rule. Consider the example in Fig.~\ref{fig:same_type_simplification}. If the nodes are in different clusters, sweeping would not contract the nodes, but would simplify all the edges except one in the same way as a the fusion rule (ignoring once again an overall factor).

Finally, the \textit{scalar decomposition} occurs when there are two nodes of the same type and at least one shares all its edges with the other. A sweep will merge the two nodes, and then only factor out a scalar (degree-$0$ node) in the decomposition to return to two tensors. However, the sweep will remove all edges between the tensors.

We now argue that these simplifications are sufficient to characterize any possible simplification present in the (2,3)-biregular graph ensemble. Each variable node has degree $2$, so the bialgebra law and any higher-order generalizations cannot occur because they require variable nodes of degree at least $3$. Because we replace any degree-$2$ variable node in a cluster by an edge and the fusion rule combines clause nodes within a cluster, most clusters will be a single clause node of some degree. Our rules above capture simplifications between such clusters. The one exception is that variable nodes are their own clusters at the start of the algorithm before being contracted with other nodes. In this case, the simplifications given by Fig.~\ref{fig:triangle_rule} may apply. Therefore, our set of graphical rules should be sufficient to capture all possible simplifications in this ensemble. We also provide evidence of this claim in Sec.~\ref{only_cores}.

\subsection{Numerical experiments and tools}
\subsubsection{Generation of random instances}
To generate our instances at a given $\alpha$ and $n$, we choose $m = \alpha n$ tuples~\footnote{Note that we choose $m,n,$ and $\alpha$ such that $m$ and $n$ are integers.} of $p$ variables uniformly at random without replacement from $\{x\}$, the set of variables defined in Sec.~\ref{pxorsat_problem}.
This means that each variable tensor's rank $d$ conforms to the following Poisson distribution:
\begin{equation}\label{eq:poisson}
    \mathcal{P}(\text{rank}(x_i) = d) = \frac{(p\alpha)^d}{d!} e^{-p\alpha}.
\end{equation}
This rank is defined as the number of times that a variable is present in the problem.
In the language of Eq.~\ref{eq:Ax=b}, we randomly place $p$ ones in each row of $A$ and the rank of the variable $x_i$ corresponds to the number of ones in column $i$. For our numerical experiments, we set $p=3$. We also exclusively focus on the case $\vec{b} = \vec{0}$ (the unfrustrated version of the $p$-spin model). We do so because in the regime $\alpha < \alpha_c$ that we study, the problem will contain at least one solution for any given $\vec{b}$ (with high probability in the limit of large problem size), which allows us to redefine the problem such that $\vec{b} = \vec{0}$~\cite{mezard_alternative_2002, braunstein_complexity_2002} and the solution count remains the same.

\subsubsection{Generation of core instances}\label{Core}
Since we are mainly concerned with the scaling of resources for instances which contain a core, we choose a minimal ensemble with this property. We will study the ensemble of connected $3$-regular graphs on $m$ clause nodes generated uniformly at random using the \texttt{Degree\_Sequence} function in \texttt{igraph} with the Viger-Latapy method~\cite{viger_efficient_2016}. To create a $3$-XORSAT instance, we place a variable node along each edge of the regular graph. This ensures the variable nodes all have degree two, and the clause nodes have degree three. Therefore, the ensemble of instances is for $\alpha = 2/3$. Note that this is below $\alpha_d$, but the method of construction explicitly ensures a core.

\subsubsection{Implementation of contraction methods}
For TN contractions, we use \texttt{quimb}, a Python package for manipulating TNs~\cite{gray2018quimb}. For the graphical method, we use \texttt{igraph}, an efficient network analysis library~\cite{csardi_igraph_nodate}, in order to work with node attributes on the graph directly. Those attributes let us define the node types (clause and variable) and the nodes' clusters.

The TN contraction order, as discussed in Sec.~\ref{sec:contraction-ordering}, determines the contraction width. Without applying our sweeping method, one can track this quantity without actually performing the tensor contraction. One must simply keep track of the ranks of the tensors at any point in the contraction, noting as in Sec.~\ref{sec:contraction} that combining two tensors yields a new tensor of known rank. We use \texttt{cotengra}, a Python package for TN contractions, to track this quantity~\cite{gray_hyper-optimized_2021}. In order to track this quantity when sweeps are applied, we use \texttt{quimb} in order to read the tensors' sizes during the contraction and calculate the contraction width using Eq.~\ref{eq:contraction-width}.

For random TNs such as ours, there exist multiple heuristic algorithms for finding contraction orderings~\cite{gray_hyper-optimized_2021, gray_hyper-optimized_2022} which lower the contraction width and are practically useful for carrying out computations. For the results in Sec.~\ref{Results}, we determine the ordering using a community detection algorithm based on the edge betweenness centrality~\cite{girvan_community_2002} (\texttt{EBC}) of the network. This algorithm is implemented as \texttt{community\_edge\_betweenness} in the Python package \texttt{igraph}~\cite{csardi_igraph_nodate}. We use the \texttt{EBC} algorithm because it looks for communities in the graph, thus contracting dense sections first. This is useful in random TNs because it minimizes the chances of having to work with huge tensors quickly, which could result in a tensor of large rank (and therefore, large contraction width). This algorithm is also deterministic, ensuring reproducibility of the contraction orderings. Furthermore, in Sec.~\ref{only_cores}, we compare the results obtained using this contraction ordering with two others: \texttt{KaHyPar}~\cite{kahypar2016, kahypar2017} and \texttt{greedy}, both from the Python package \texttt{cotengra}.

Even with these better contraction orderings, exactly contracting these random TNs without bond compression will generally result in an exponential growth in $n$ of time and memory (see Sec.~\ref{Results}). However, we will show that by manipulating the TN after each contraction using the algorithm defined in Sec.~\ref{sec:sweeping-method}, we can alter the scaling of resources for a range of parameter values in the problem.

\subsubsection{Sweeping Method} \label{sec:sweeping-method}
To ensure lossless compression in bond sweeping, we set the relative threshold for zero singular values to be $10^{-12}$. We sweep the TN in arbitrary order until the tensor sizes converge. During a sweep, we compress all the bonds using the \texttt{compress\_all} method implemented in \texttt{quimb}, which uses the compression schedule described in Eq.~\ref{eq:compression_schedule}. We perform sweeps before each contraction, potentially finding simplifications (see Sec.~\ref{sec:XORSAT-simplifications}) in the structure of the TN during each step of the full contraction.

\section{Results}\label{Results}

\subsection{Numerical contraction for random instances}\label{Results_tensors}
\begin{figure}[htbp]
    \begin{center}
    \includegraphics[width=0.5\textwidth]{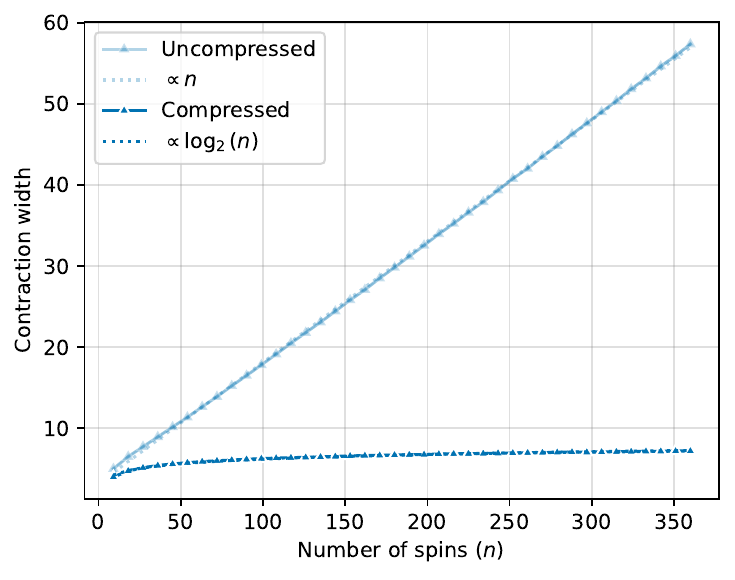}
    \end{center}
    \caption{Average contraction width for $\alpha=2/3$ without (light) and with (dark) compression.}
    \label{fig:alpha066_both_memory_curves}
\end{figure}

Numerical TN contractions were performed on an AMD EPYC 7F72 @ 3.2 GHz processor, with a maximum allocated RAM of 1 TB.
Each point in the figures of this section corresponds to the median contraction width or contraction runtime over $10^4$ instances for a given number of spins $n$, except for Fig.~\ref{fig:alpha066_both_memory_curves} which shows the average scaling of the contraction width.
The contraction width determines, to leading order, the contraction runtime. As is common in random graph ensembles for spin-glass models or Boolean variable graphical models, the instance samples contains outliers that are much harder to solve than the typical instance.

In Fig.~\ref{fig:alpha066_both_memory_curves}, we show the average contraction width with and without compression (sweeping) for $\alpha = 2/3$. Without compression, the scaling of the average contraction width is linear, indicating exponential growth of tensor sizes. By contrast, compression changes the scaling to one that is well described by a logarithmic curve, indicating polynomially growing tensor sizes and hence contraction runtimes.

\begin{figure*}[htbp]
    \begin{center}
    \includegraphics[width=\textwidth]{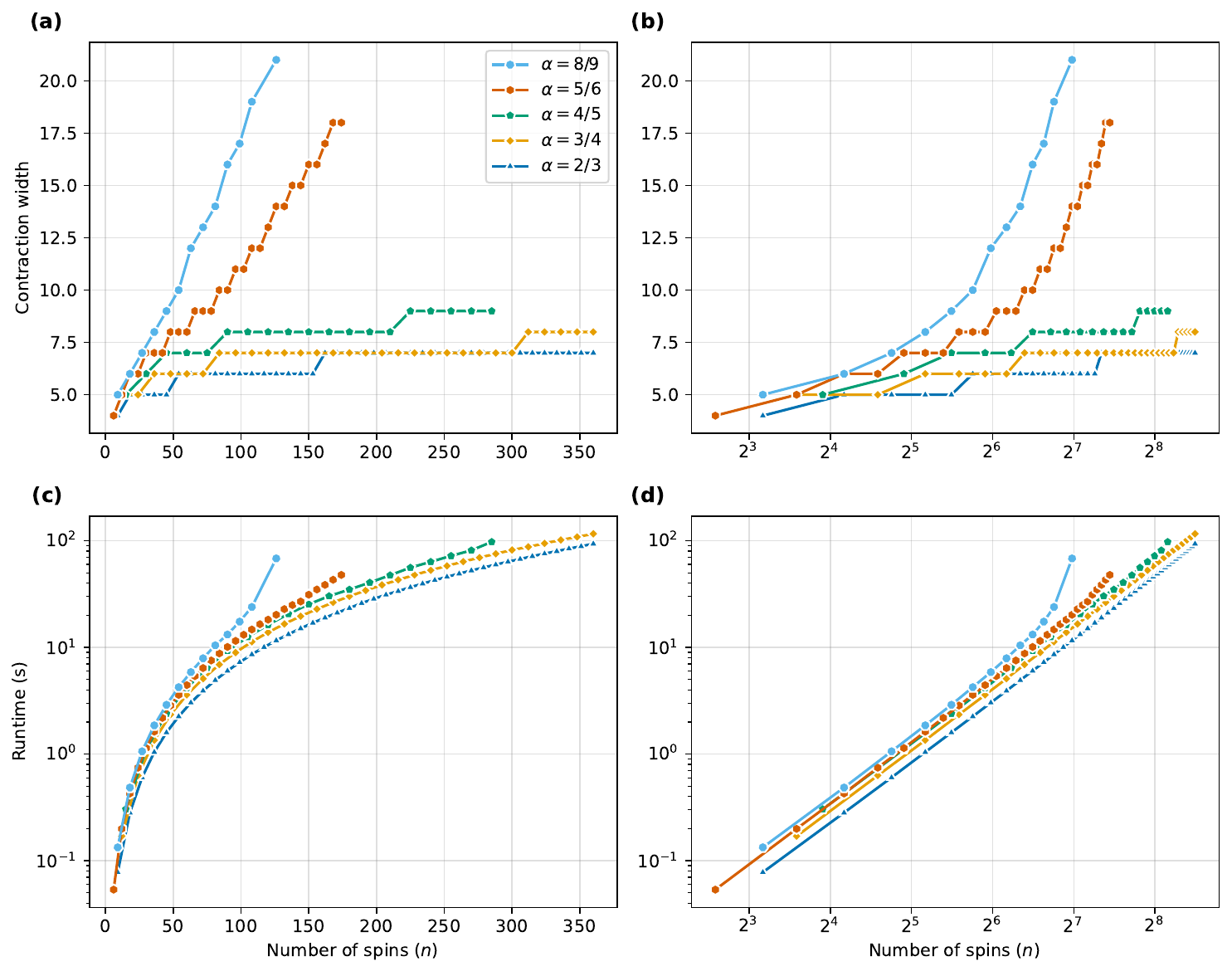}
    \end{center}
    \caption{Scaling of the contraction width and runtime of compressed TN contraction for the $3$-spin model. \textbf{(a)} The median contraction width. \textbf{(b)} The same data as in (a), but on a logarithmic horizontal scale to accentuate the curves which follow a logarithmic scale (which can be fitted with straight lines). \textbf{(c)} Our algorithm's compressed contraction median runtime. This panel shows the exponential scaling by straight lines. \textbf{(d)} The same data as in (c), but shown on a horizontal logarithmic scale to accentuate the curves which follow polynomial scaling (straight lines).}
    \label{fig:all_quimb_results}
\end{figure*}

We studied larger values of $\alpha$ and we show in Fig.~\ref{fig:all_quimb_results} how the scaling of both the median contraction width and median contraction runtime evolve as $\alpha$ increases.
For the largest system sizes, out of $10^4$ instances, a few outliers require times beyond any reasonable timeout we have tried, as expected.
We therefore cannot report unbiased runtime averages for these sizes.
However, when plotted against system size, the data for the average and median of the contraction width are comparable (likewise for the contraction runtime). Note that we observe that the average contraction width is a smoother function of system size than the median contraction width; though we show the median contraction width in Fig.~\ref{fig:all_quimb_results}, we use the average contraction width data to extract a scaling. We do the same for the contraction runtime. In this case, the curves in the bottom of Fig.~\ref{fig:all_quimb_results} are already smooth.

The results in Fig.~\ref{fig:all_quimb_results}(a) highlight linear scaling of the curves for $\alpha = 5/6$ and $\alpha=8/9$ while Fig.~\ref{fig:all_quimb_results}(b) clearly shows the logarithmic nature of the curves for $\alpha = 2/3$ and $\alpha = 3/4$. For $\alpha = 4/5$, this median scaling seems to be of logarithmic nature in Fig.~\ref{fig:all_quimb_results}(b), but analysing the average shows that it actually starts to ``peel-off'' from logarithmic scaling.

The logarithmic scaling for $\alpha = 2/3$ and $\alpha = 3/4$ is mainly due to the TN algorithm automatically implementing leaf removal, since $\alpha < \alpha_d$.
Indeed, this leads to a high probability that the initial sweeps will remove all the edges in the TN even before the first contraction, leaving only scalars to be multiplied.
For $\alpha = 5/6$ and $\alpha = 8/9$, values that are greater than $\alpha_d$, we find that the algorithm is less efficient due to a core that remains after the initial sweeps.
Those cores lead to actual tensor contractions instead of scalar multiplications, so the instances with $\alpha = 5/6$ and $\alpha = 8/9$ become harder to compute, hence the contraction widths' polynomial scaling.
As we noted, since $\alpha = 4/5$ is close to $\alpha_d$, there is a probability of a core remaining for our finite system sizes, so the algorithm starts becoming less efficient here too.
Sweeping still removes all the edges in the TN in most cases, but less so than with $\alpha = 3/4$, thus the ``peel-off'' starting at $\alpha = 4/5$.

In Fig.~\ref{fig:all_quimb_results}(c), we see the scaling of the median contraction runtime (in seconds) with a logarithmic vertical axis and the same data is shown with a logarithmic horizontal axis in Fig.~\ref{fig:all_quimb_results}(d).
Accordingly with the contraction width scaling, we find polynomial curves for $\alpha = 2/3$ ($\propto n^{1.928}$) and $\alpha=3/4$ ($\propto n^{1.902}$).
For smaller $n$, we see that the time scaling for all the curves follow a polynomial scaling.
This is due to the small finite size of the TN, since it changes for bigger TNs, or for larger $n$.
The ``peel-off'' phenomenon is thus also observed at the end of the curves for $\alpha \in \left\{4/5, 5/6, 8/9 \right\}$, becoming more pronounced with increasing $\alpha$.
This means that the scaling transitions from polynomial to superpolynomial, like the conclusion on memory usage in Fig.~\ref{fig:all_quimb_results}(b).

At $\alpha = 2/3$ and $3/4$, the compressed TN algorithm exhibits performance between those of the standard and ``smart'' GE methods (see Table~\ref{tab:scaling_comparisons}). For these values of the $\alpha$ parameter, the contraction runtime can be further improved by removing bonds of dimension $1$ after each contraction step. Indeed, when a bond is completely compressed by our algorithm, a dimension $1$ bond remains between the two neighboring tensors. These dimension $1$ bonds do not affect memory scaling, yet the sweeping algorithm will continue trying to compress them, even though they cannot be further compressed. Eliminating those ``useless'' bonds results in improved polynomial contraction runtimes for $\alpha = 2/3$ and $3/4$, as shown in Table~\ref{tab:scaling_comparisons}, since the subsequent sweeps will not try to compress those bonds anymore. In this same table, the memory is defined as the maximum size of the whole TN --- the sum of all its tensors' sizes --- reached during its contraction with the sweeping method applied.

\begin{table}[htbp]
    \centering
    \begin{tabular}{|c|c|c|c|}
        \hline
        \text{Methods} & $\alpha$ & \text{Memory} & \text{Time}\\
        \hline\hline
        \text{Standard GE} & $2/3$ & $\propto n$ & $\propto n$\\
        \hline
        \text{Standard GE} & $3/4$ & $\propto n^2$ & $\propto n^3$\\
        \hline
        \text{Smart GE} & $ < \alpha_\mathrm{d}$ & $\propto n$ & $\propto n$\\
        \hline
        \text{Compressed TN} & $2/3$ & $\propto n^{1.030}$ & $\propto n^{1.226}$\\
        \hline
        \text{Compressed TN} & $3/4$ & $\propto n^{1.112}$ & $\propto n^{1.420}$\\
        \hline
    \end{tabular}
    \caption{Performance comparison between optimized compressed TN contraction and GE.}
    \label{tab:scaling_comparisons}
\end{table}

\subsection{Graphical contraction for leaf-free instances}\label{only_cores}
\begin{figure}[htbp]
    \begin{center}
    \includegraphics[width=0.5\textwidth]{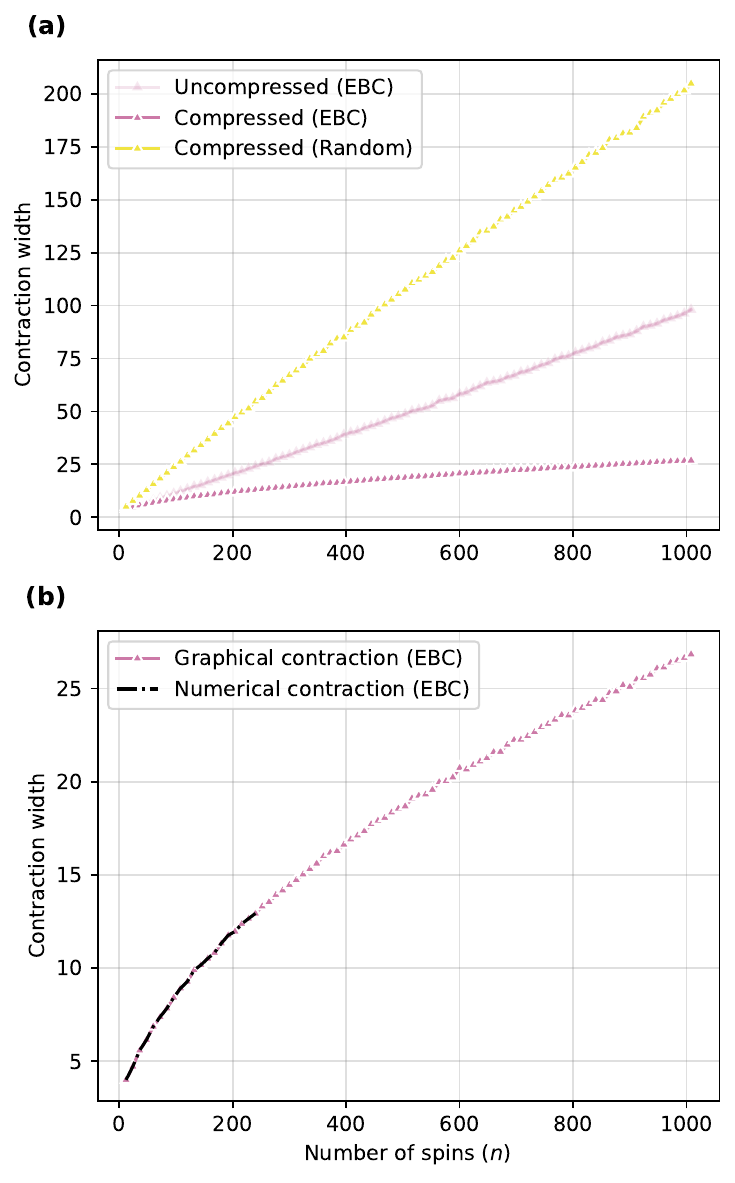}
    \end{center}
    \caption{\textbf{(a)} Scaling of the average contraction width for instances in the (2,3)-biregular graph ensemble ($\alpha = 2/3$ leaf-free instances) using the \texttt{EBC} contraction ordering with and without sweeping, and using the \texttt{Random} contraction ordering with sweeping. We obtained all results using graphical contractions. \textbf{(b)} Comparison of the scaling of the average contraction width (with the \texttt{EBC} contraction ordering) obtained using graphical contractions and obtained using the numerical contractions with sweeps applied.}
    \label{fig:results_CEB_graphical}
\end{figure}

\begin{figure}
    \centering
    \includegraphics[width=0.5\textwidth]{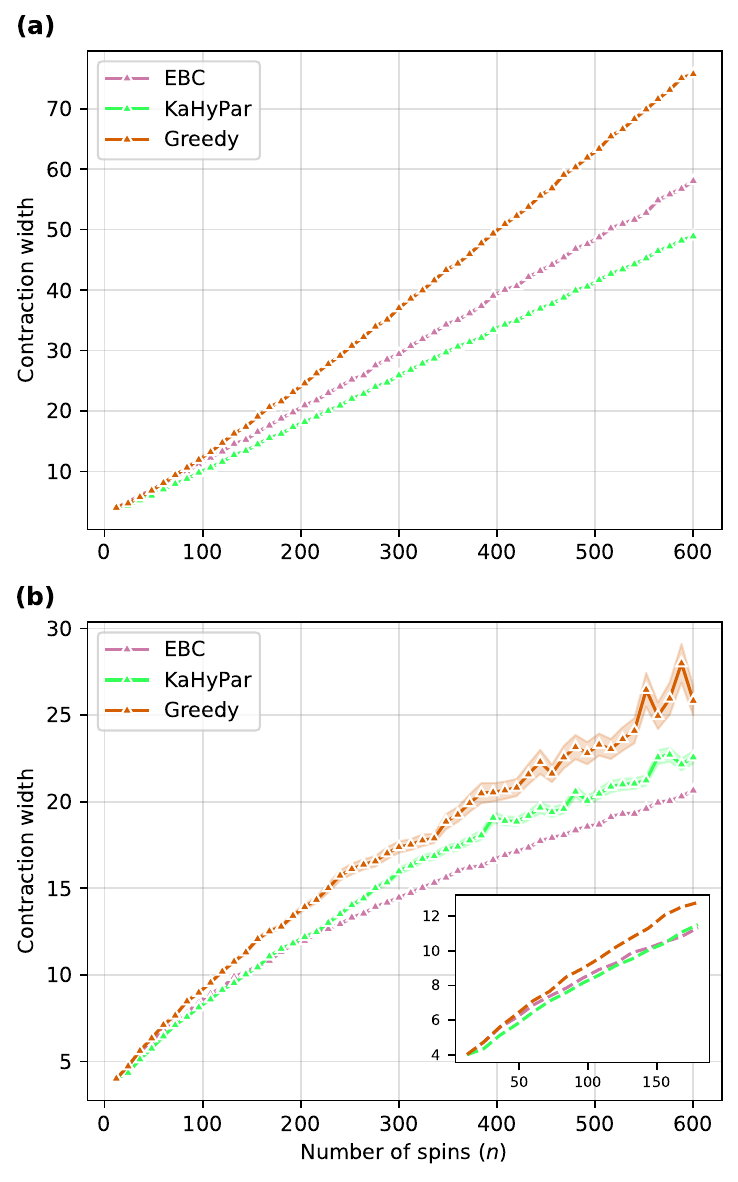}
    \caption{All the results were obtained using graphical contractions. \textbf{(a)} Scaling of the average contraction width for instances in the (2,3)-biregular graph ensemble ($\alpha = 2/3$ leaf-free instances) using the \texttt{EBC}, \texttt{KaHyPar} and \texttt{greedy} contraction orderings without sweeping. \textbf{(b)} Comparison of the average contraction width for the same ensemble using the same three contraction orderings, but with sweeps applied.}
    \label{fig:kahypar-vs-greedy-vs-ebc}
\end{figure}

For the leaf-free ensemble, each point in the figures has been averaged over $200$ random leaf-free instances.
With the graphical method, the contraction widths are extracted from the number of clusters' outgoing edges during the TN contraction, as explained in Sec.~\ref{sec:graphical-method}.
All the results for the contraction width obtained with this graphical method are shown in Fig.~\ref{fig:results_CEB_graphical} and Fig.~\ref{fig:kahypar-vs-greedy-vs-ebc}.

Now having the possibility to study larger TNs without being limited by the memory, we can compare the contraction width of the algorithm on different contraction orderings. In Fig.~\ref{fig:results_CEB_graphical}(a), we compare two of them: \texttt{EBC} and \texttt{Random}. The \texttt{Random} method chooses the next tensors to be contracted completely randomly. It can thus only be usefully studied with this graphical method because it quickly scales to astronomical contraction widths, as seen in Fig.~\ref{fig:results_CEB_graphical}(a).

From Fig.~\ref{fig:results_CEB_graphical}(a), we see that a good contraction ordering is an important factor for the success of the sweeping method during the contraction of a given TN that models a $p$-XORSAT problem. Two known methods for random tensor networks have also been used in order to compare the results obtained from the \texttt{EBC} method, as shown in Fig.~\ref{fig:kahypar-vs-greedy-vs-ebc}.

The results demonstrate that the sweeping method finds enough simplifications for instances in the (2,3)-biregular graph ensemble so that the scaling of the average contraction width changes from linear to sublinear for the \texttt{EBC}, \texttt{KaHyPar} and \texttt{greedy} contraction orderings.
From those results, after $n \approx 150$, we see that the \texttt{EBC} method is most efficient in finding those simplifications of the three, followed by \texttt{KaHyPar} and then \texttt{greedy}.
The precise functional form of the scaling is nontrivial and we have not been able to determine a sufficiently accurate fitting function. This means that the sweeping method goes beyond the efficacy of the leaf removal in the TN representation of the $3$-spin model.

Note that the (2,3)-biregular graph ensemble we consider offers a simplification of the corresponding $p$-XORSAT problem which allows leaf removal --- and by extension, our TN algorithm --- to work efficiently in polynomial time. Suppose $A$ is the $m \times n$ matrix encoding the problem. By definition of the ensemble, each column of $A$ has exactly two $1$s. Therefore, the rows of $A$ satisfy $A_1 + A_2 + \cdots + A_m = 0 \mod 2$, indicating the rows are linearly dependent. In other words, we can remove some row $A_i$ from the problem without changing the solution space and count. In terms of the graph, one can remove the corresponding clause node encoding $A_i$ because it is made redundant by the other clauses. However, removing a clause node allows leaf removal to begin since the variable nodes that were connected to that removed clause node will now be degree-$1$. Leaf removal will then succeed in solving the problem and producing an empty core, implying both leaf removal and our TN algorithm are efficient for this ensemble if we first remove a single redundant clause.

We have verified that the graphical contraction method of Sec.~\ref{sec:graphical-method} yields tensor sizes identical to those found via numerical contraction \emph{at each contraction step} by comparing the two methods for 100 random instances with $n = 81$ (for the \texttt{EBC} and \texttt{Random} contraction orderings). Moreover, all contraction widths for the 200 random instances used to get the results in Fig.~\ref{fig:results_CEB_graphical}(b) with sizes up to $n = 240$ are identical to those obtained with numerical contraction (for the \texttt{EBC} contraction ordering).

\section{Conclusion} \label{sec:Conclusion}
In this work, we have applied compressed TN contraction to the $p$-spin model. Focusing on $p=3$, we have shown that lossless compression sweeps over the bonds of the network emulate the leaf removal algorithm, meaning that the TN method is efficient (i.e., polynomial-time) below the dynamical transition at $\alpha_d \approx 0.818$. Above the dynamical transition, the appearance of a leafless core adversely affects the performance of the TN algorithm, which is now superpolynomial-time. Nevertheless, by focusing on the restricted ensemble of biregular instances where every spin participates in exactly two interactions, we find that compressed contraction can be done in \emph{subexponential} time. This speedup over the anticipated exponential scaling depends crucially on the choice of contraction path. We note that, unlike some previous TN techniques applied to spin-glass models~\cite{zhu2019tensor}, our methods are exact and can be made to suffer no loss of precision for the case of XOR constraints. Indeed, we observe that the local singular values during each sweeping step correspond to either positive or fractional powers of two if they are distributed properly after having applied the SVD. This means that we either have those values or zero/numerical zero singular values. A similar observation has been made for Clifford circuits, essentially parity circuits, where stabilizer states possess flat entanglement spectra~\cite{fattal2004entanglement, Hamma2005entropy, zhou2020Clifford}. To our knowledge, this is the first general-purpose numerical method for spin-model partition function and model counting computations that achieves this performance for $p$-spin models without invoking GE as a subroutine. Furthermore, we believe that this is the first nontrivial case of a spin model defined on random sparse graphs (that are \emph{not} trees) where compressed TN contraction solves the model exactly, yet leads to an exponential-to-subexponential speedup over direct TN contraction.

\section{Acknowledgements}
This work was supported by the Minist\`{e}re de l'\'{E}conomie, de l'Innovation et de l'Énergie du Qu\'{e}bec through its Research Chair in Quantum Computing, an NSERC Discovery grant, and the Canada First Research Excellence Fund. This work made use of the compute infrastructure of Calcul Québec and the Digital Research Alliance of Canada.

\bibliographystyle{apsrev4-1}
\bibliography{main}

\begin{thebibliography}{48}%
\makeatletter
\providecommand \@ifxundefined [1]{%
 \@ifx{#1\undefined}
}%
\providecommand \@ifnum [1]{%
 \ifnum #1\expandafter \@firstoftwo
 \else \expandafter \@secondoftwo
 \fi
}%
\providecommand \@ifx [1]{%
 \ifx #1\expandafter \@firstoftwo
 \else \expandafter \@secondoftwo
 \fi
}%
\providecommand \natexlab [1]{#1}%
\providecommand \enquote  [1]{``#1''}%
\providecommand \bibnamefont  [1]{#1}%
\providecommand \bibfnamefont [1]{#1}%
\providecommand \citenamefont [1]{#1}%
\providecommand \href@noop [0]{\@secondoftwo}%
\providecommand \href [0]{\begingroup \@sanitize@url \@href}%
\providecommand \@href[1]{\@@startlink{#1}\@@href}%
\providecommand \@@href[1]{\endgroup#1\@@endlink}%
\providecommand \@sanitize@url [0]{\catcode `\\12\catcode `\$12\catcode `\&12\catcode `\#12\catcode `\^12\catcode `\_12\catcode `\%12\relax}%
\providecommand \@@startlink[1]{}%
\providecommand \@@endlink[0]{}%
\providecommand \url  [0]{\begingroup\@sanitize@url \@url }%
\providecommand \@url [1]{\endgroup\@href {#1}{\urlprefix }}%
\providecommand \urlprefix  [0]{URL }%
\providecommand \Eprint [0]{\href }%
\providecommand \doibase [0]{http://dx.doi.org/}%
\providecommand \selectlanguage [0]{\@gobble}%
\providecommand \bibinfo  [0]{\@secondoftwo}%
\providecommand \bibfield  [0]{\@secondoftwo}%
\providecommand \translation [1]{[#1]}%
\providecommand \BibitemOpen [0]{}%
\providecommand \bibitemStop [0]{}%
\providecommand \bibitemNoStop [0]{.\EOS\space}%
\providecommand \EOS [0]{\spacefactor3000\relax}%
\providecommand \BibitemShut  [1]{\csname bibitem#1\endcsname}%
\let\auto@bib@innerbib\@empty
\bibitem [{\citenamefont {Kirkpatrick}\ and\ \citenamefont {Toulouse}(1985)}]{kirkpatrick1985configuration}%
  \BibitemOpen
  \bibfield  {author} {\bibinfo {author} {\bibfnamefont {S.}~\bibnamefont {Kirkpatrick}}\ and\ \bibinfo {author} {\bibfnamefont {G.}~\bibnamefont {Toulouse}},\ }\href {\doibase 10.1051/jphys:019850046080127700} {\bibfield  {journal} {\bibinfo  {journal} {J. Phys. France}\ }\textbf {\bibinfo {volume} {46}},\ \bibinfo {pages} {1277} (\bibinfo {year} {1985})}\BibitemShut {NoStop}%
\bibitem [{\citenamefont {Bryngelson}\ and\ \citenamefont {Wolynes}(1987)}]{bryngelson1987spin}%
  \BibitemOpen
  \bibfield  {author} {\bibinfo {author} {\bibfnamefont {J.~D.}\ \bibnamefont {Bryngelson}}\ and\ \bibinfo {author} {\bibfnamefont {P.~G.}\ \bibnamefont {Wolynes}},\ }\href {\doibase 10.1073/pnas.84.21.7524} {\bibfield  {journal} {\bibinfo  {journal} {Proc. Natl. Acad. Sci.}\ }\textbf {\bibinfo {volume} {84}},\ \bibinfo {pages} {7524} (\bibinfo {year} {1987})}\BibitemShut {NoStop}%
\bibitem [{\citenamefont {Venkataraman}\ and\ \citenamefont {Athithan}(1991)}]{venkataraman1991spin}%
  \BibitemOpen
  \bibfield  {author} {\bibinfo {author} {\bibfnamefont {G.}~\bibnamefont {Venkataraman}}\ and\ \bibinfo {author} {\bibfnamefont {G.}~\bibnamefont {Athithan}},\ }\href {\doibase 10.1007/BF02846491} {\bibfield  {journal} {\bibinfo  {journal} {Pramana}\ }\textbf {\bibinfo {volume} {36}},\ \bibinfo {pages} {1} (\bibinfo {year} {1991})}\BibitemShut {NoStop}%
\bibitem [{\citenamefont {Stein}\ and\ \citenamefont {Newman}(2013)}]{stein2013spin}%
  \BibitemOpen
  \bibfield  {author} {\bibinfo {author} {\bibfnamefont {D.~L.}\ \bibnamefont {Stein}}\ and\ \bibinfo {author} {\bibfnamefont {C.~M.}\ \bibnamefont {Newman}},\ }\href {\doibase 10.23943/princeton/9780691147338.001.0001} {\emph {\bibinfo {title} {{S}pin {G}lasses and {C}omplexity}}}\ (\bibinfo  {publisher} {Princeton University Press},\ \bibinfo {year} {2013})\BibitemShut {NoStop}%
\bibitem [{\citenamefont {M\'{e}zard}\ \emph {et~al.}(2003)\citenamefont {M\'{e}zard}, \citenamefont {Ricci-Tersenghi},\ and\ \citenamefont {Zecchina}}]{mezard_alternative_2002}%
  \BibitemOpen
  \bibfield  {author} {\bibinfo {author} {\bibfnamefont {M.}~\bibnamefont {M\'{e}zard}}, \bibinfo {author} {\bibfnamefont {F.}~\bibnamefont {Ricci-Tersenghi}}, \ and\ \bibinfo {author} {\bibfnamefont {R.}~\bibnamefont {Zecchina}},\ }\href {\doibase 10.1023/A:1022886412117} {\bibfield  {journal} {\bibinfo  {journal} {J. Stat. Phys.}\ } (\bibinfo {year} {2003}),\ 10.1023/A:1022886412117}\BibitemShut {NoStop}%
\bibitem [{\citenamefont {Ricci-Tersenghi}(2010)}]{ricci-tersenghi_being_2010}%
  \BibitemOpen
  \bibfield  {author} {\bibinfo {author} {\bibfnamefont {F.}~\bibnamefont {Ricci-Tersenghi}},\ }\href {\doibase 10.1126/science.1189804} {\bibfield  {journal} {\bibinfo  {journal} {Science}\ }\textbf {\bibinfo {volume} {330}},\ \bibinfo {pages} {1639} (\bibinfo {year} {2010})}\BibitemShut {NoStop}%
\bibitem [{\citenamefont {Bernaschi}\ \emph {et~al.}(2021)\citenamefont {Bernaschi}, \citenamefont {Bisson}, \citenamefont {Fatica}, \citenamefont {Marinari}, \citenamefont {Martin-Mayor}, \citenamefont {Parisi},\ and\ \citenamefont {Ricci-Tersenghi}}]{bernaschi2021we}%
  \BibitemOpen
  \bibfield  {author} {\bibinfo {author} {\bibfnamefont {M.}~\bibnamefont {Bernaschi}}, \bibinfo {author} {\bibfnamefont {M.}~\bibnamefont {Bisson}}, \bibinfo {author} {\bibfnamefont {M.}~\bibnamefont {Fatica}}, \bibinfo {author} {\bibfnamefont {E.}~\bibnamefont {Marinari}}, \bibinfo {author} {\bibfnamefont {V.}~\bibnamefont {Martin-Mayor}}, \bibinfo {author} {\bibfnamefont {G.}~\bibnamefont {Parisi}}, \ and\ \bibinfo {author} {\bibfnamefont {F.}~\bibnamefont {Ricci-Tersenghi}},\ }\href {\doibase 10.1209/0295-5075/133/60005} {\bibfield  {journal} {\bibinfo  {journal} {Europhysics Letters}\ }\textbf {\bibinfo {volume} {133}},\ \bibinfo {pages} {60005} (\bibinfo {year} {2021})}\BibitemShut {NoStop}%
\bibitem [{\citenamefont {Kanao}\ and\ \citenamefont {Goto}(2022)}]{kanao2022simulated}%
  \BibitemOpen
  \bibfield  {author} {\bibinfo {author} {\bibfnamefont {T.}~\bibnamefont {Kanao}}\ and\ \bibinfo {author} {\bibfnamefont {H.}~\bibnamefont {Goto}},\ }\href {\doibase 10.35848/1882-0786/acaba9} {\bibfield  {journal} {\bibinfo  {journal} {Applied Physics Express}\ }\textbf {\bibinfo {volume} {16}},\ \bibinfo {pages} {014501} (\bibinfo {year} {2022})}\BibitemShut {NoStop}%
\bibitem [{\citenamefont {Aadit}\ \emph {et~al.}(2023)\citenamefont {Aadit}, \citenamefont {Nikhar}, \citenamefont {Kannan}, \citenamefont {Chowdhury},\ and\ \citenamefont {Camsari}}]{aadit2023all}%
  \BibitemOpen
  \bibfield  {author} {\bibinfo {author} {\bibfnamefont {N.~A.}\ \bibnamefont {Aadit}}, \bibinfo {author} {\bibfnamefont {S.}~\bibnamefont {Nikhar}}, \bibinfo {author} {\bibfnamefont {S.}~\bibnamefont {Kannan}}, \bibinfo {author} {\bibfnamefont {S.}~\bibnamefont {Chowdhury}}, \ and\ \bibinfo {author} {\bibfnamefont {K.~Y.}\ \bibnamefont {Camsari}},\ }\href {\doibase 2312.08748} {\bibfield  {journal} {\bibinfo  {journal} {arXiv preprint arXiv:2312.08748}\ } (\bibinfo {year} {2023}),\ 2312.08748}\BibitemShut {NoStop}%
\bibitem [{\citenamefont {J\"org}\ \emph {et~al.}(2010)\citenamefont {J\"org}, \citenamefont {Krzakala}, \citenamefont {Semerjian},\ and\ \citenamefont {Zamponi}}]{jorg2010first}%
  \BibitemOpen
  \bibfield  {author} {\bibinfo {author} {\bibfnamefont {T.}~\bibnamefont {J\"org}}, \bibinfo {author} {\bibfnamefont {F.}~\bibnamefont {Krzakala}}, \bibinfo {author} {\bibfnamefont {G.}~\bibnamefont {Semerjian}}, \ and\ \bibinfo {author} {\bibfnamefont {F.}~\bibnamefont {Zamponi}},\ }\href {\doibase 10.1103/PhysRevLett.104.207206} {\bibfield  {journal} {\bibinfo  {journal} {Phys. Rev. Lett.}\ }\textbf {\bibinfo {volume} {104}},\ \bibinfo {pages} {207206} (\bibinfo {year} {2010})}\BibitemShut {NoStop}%
\bibitem [{\citenamefont {Farhi}\ \emph {et~al.}(2012)\citenamefont {Farhi}, \citenamefont {Gosset}, \citenamefont {Hen}, \citenamefont {Sandvik}, \citenamefont {Shor}, \citenamefont {Young},\ and\ \citenamefont {Zamponi}}]{farhi2012performance}%
  \BibitemOpen
  \bibfield  {author} {\bibinfo {author} {\bibfnamefont {E.}~\bibnamefont {Farhi}}, \bibinfo {author} {\bibfnamefont {D.}~\bibnamefont {Gosset}}, \bibinfo {author} {\bibfnamefont {I.}~\bibnamefont {Hen}}, \bibinfo {author} {\bibfnamefont {A.~W.}\ \bibnamefont {Sandvik}}, \bibinfo {author} {\bibfnamefont {P.}~\bibnamefont {Shor}}, \bibinfo {author} {\bibfnamefont {A.~P.}\ \bibnamefont {Young}}, \ and\ \bibinfo {author} {\bibfnamefont {F.}~\bibnamefont {Zamponi}},\ }\href {\doibase 10.1103/PhysRevA.86.052334} {\bibfield  {journal} {\bibinfo  {journal} {Phys. Rev. A}\ }\textbf {\bibinfo {volume} {86}},\ \bibinfo {pages} {052334} (\bibinfo {year} {2012})}\BibitemShut {NoStop}%
\bibitem [{\citenamefont {Hen}(2019)}]{hen2019equation}%
  \BibitemOpen
  \bibfield  {author} {\bibinfo {author} {\bibfnamefont {I.}~\bibnamefont {Hen}},\ }\href {\doibase 10.1103/PhysRevApplied.12.011003} {\bibfield  {journal} {\bibinfo  {journal} {Phys. Rev. Appl.}\ }\textbf {\bibinfo {volume} {12}},\ \bibinfo {pages} {011003} (\bibinfo {year} {2019})}\BibitemShut {NoStop}%
\bibitem [{\citenamefont {Bellitti}\ \emph {et~al.}(2021)\citenamefont {Bellitti}, \citenamefont {Ricci-Tersenghi},\ and\ \citenamefont {Scardicchio}}]{bellitti2021entropic}%
  \BibitemOpen
  \bibfield  {author} {\bibinfo {author} {\bibfnamefont {M.}~\bibnamefont {Bellitti}}, \bibinfo {author} {\bibfnamefont {F.}~\bibnamefont {Ricci-Tersenghi}}, \ and\ \bibinfo {author} {\bibfnamefont {A.}~\bibnamefont {Scardicchio}},\ }\href {\doibase 10.1103/PhysRevResearch.3.043015} {\bibfield  {journal} {\bibinfo  {journal} {Phys. Rev. Res.}\ }\textbf {\bibinfo {volume} {3}},\ \bibinfo {pages} {043015} (\bibinfo {year} {2021})}\BibitemShut {NoStop}%
\bibitem [{\citenamefont {Kowalsky}\ \emph {et~al.}(2022)\citenamefont {Kowalsky}, \citenamefont {Albash}, \citenamefont {Hen},\ and\ \citenamefont {Lidar}}]{kowalsky20223}%
  \BibitemOpen
  \bibfield  {author} {\bibinfo {author} {\bibfnamefont {M.}~\bibnamefont {Kowalsky}}, \bibinfo {author} {\bibfnamefont {T.}~\bibnamefont {Albash}}, \bibinfo {author} {\bibfnamefont {I.}~\bibnamefont {Hen}}, \ and\ \bibinfo {author} {\bibfnamefont {D.~A.}\ \bibnamefont {Lidar}},\ }\href {\doibase 10.1088/2058-9565/ac4d1b} {\bibfield  {journal} {\bibinfo  {journal} {Quantum Science and Technology}\ }\textbf {\bibinfo {volume} {7}},\ \bibinfo {pages} {025008} (\bibinfo {year} {2022})}\BibitemShut {NoStop}%
\bibitem [{\citenamefont {Patil}\ \emph {et~al.}(2019)\citenamefont {Patil}, \citenamefont {Kourtis}, \citenamefont {Chamon}, \citenamefont {Mucciolo},\ and\ \citenamefont {Ruckenstein}}]{patil_obstacles_2019}%
  \BibitemOpen
  \bibfield  {author} {\bibinfo {author} {\bibfnamefont {P.}~\bibnamefont {Patil}}, \bibinfo {author} {\bibfnamefont {S.}~\bibnamefont {Kourtis}}, \bibinfo {author} {\bibfnamefont {C.}~\bibnamefont {Chamon}}, \bibinfo {author} {\bibfnamefont {E.~R.}\ \bibnamefont {Mucciolo}}, \ and\ \bibinfo {author} {\bibfnamefont {A.~E.}\ \bibnamefont {Ruckenstein}},\ }\href {\doibase 10.1103/PhysRevB.100.054435} {\bibfield  {journal} {\bibinfo  {journal} {Phys. Rev. B}\ }\textbf {\bibinfo {volume} {100}},\ \bibinfo {pages} {054435} (\bibinfo {year} {2019})}\BibitemShut {NoStop}%
\bibitem [{\citenamefont {Haanp{\"a}{\"a}}\ \emph {et~al.}(2006)\citenamefont {Haanp{\"a}{\"a}}, \citenamefont {J{\"a}rvisalo}, \citenamefont {Kaski},\ and\ \citenamefont {Niemel{\"a}}}]{haanpaa2006hard}%
  \BibitemOpen
  \bibfield  {author} {\bibinfo {author} {\bibfnamefont {H.}~\bibnamefont {Haanp{\"a}{\"a}}}, \bibinfo {author} {\bibfnamefont {M.}~\bibnamefont {J{\"a}rvisalo}}, \bibinfo {author} {\bibfnamefont {P.}~\bibnamefont {Kaski}}, \ and\ \bibinfo {author} {\bibfnamefont {I.}~\bibnamefont {Niemel{\"a}}},\ }\href {\doibase 10.3233/SAT190015} {\bibfield  {journal} {\bibinfo  {journal} {Journal on Satisfiability, Boolean Modeling and Computation}\ }\textbf {\bibinfo {volume} {2}},\ \bibinfo {pages} {27} (\bibinfo {year} {2006})}\BibitemShut {NoStop}%
\bibitem [{\citenamefont {J{\"a}rvisalo}(2006)}]{jarvisalo2006further}%
  \BibitemOpen
  \bibfield  {author} {\bibinfo {author} {\bibfnamefont {M.}~\bibnamefont {J{\"a}rvisalo}},\ }in\ \href@noop {} {\emph {\bibinfo {booktitle} {AAAI}}}\ (\bibinfo {year} {2006})\ pp.\ \bibinfo {pages} {1873--1874}\BibitemShut {NoStop}%
\bibitem [{\citenamefont {Jia}\ \emph {et~al.}(2005)\citenamefont {Jia}, \citenamefont {Moore},\ and\ \citenamefont {Selman}}]{jia2004spin}%
  \BibitemOpen
  \bibfield  {author} {\bibinfo {author} {\bibfnamefont {H.}~\bibnamefont {Jia}}, \bibinfo {author} {\bibfnamefont {C.}~\bibnamefont {Moore}}, \ and\ \bibinfo {author} {\bibfnamefont {B.}~\bibnamefont {Selman}},\ }in\ \href@noop {} {\emph {\bibinfo {booktitle} {Theory and Applications of Satisfiability Testing}}},\ \bibinfo {editor} {edited by\ \bibinfo {editor} {\bibfnamefont {H.~H.}\ \bibnamefont {Hoos}}\ and\ \bibinfo {editor} {\bibfnamefont {D.~G.}\ \bibnamefont {Mitchell}}}\ (\bibinfo  {publisher} {Springer Berlin Heidelberg},\ \bibinfo {address} {Berlin, Heidelberg},\ \bibinfo {year} {2005})\ pp.\ \bibinfo {pages} {199--210}\BibitemShut {NoStop}%
\bibitem [{\citenamefont {Barthel}\ \emph {et~al.}(2002)\citenamefont {Barthel}, \citenamefont {Hartmann}, \citenamefont {Leone}, \citenamefont {Ricci-Tersenghi}, \citenamefont {Weigt},\ and\ \citenamefont {Zecchina}}]{barthel2002hiding}%
  \BibitemOpen
  \bibfield  {author} {\bibinfo {author} {\bibfnamefont {W.}~\bibnamefont {Barthel}}, \bibinfo {author} {\bibfnamefont {A.~K.}\ \bibnamefont {Hartmann}}, \bibinfo {author} {\bibfnamefont {M.}~\bibnamefont {Leone}}, \bibinfo {author} {\bibfnamefont {F.}~\bibnamefont {Ricci-Tersenghi}}, \bibinfo {author} {\bibfnamefont {M.}~\bibnamefont {Weigt}}, \ and\ \bibinfo {author} {\bibfnamefont {R.}~\bibnamefont {Zecchina}},\ }\href {\doibase 10.1103/PhysRevLett.88.188701} {\bibfield  {journal} {\bibinfo  {journal} {Physical Review Letter}\ }\textbf {\bibinfo {volume} {88}},\ \bibinfo {pages} {188701} (\bibinfo {year} {2002})}\BibitemShut {NoStop}%
\bibitem [{\citenamefont {Ricci-Tersenghi}\ \emph {et~al.}(2001)\citenamefont {Ricci-Tersenghi}, \citenamefont {Weigt},\ and\ \citenamefont {Zecchina}}]{ricci2001simplest}%
  \BibitemOpen
  \bibfield  {author} {\bibinfo {author} {\bibfnamefont {F.}~\bibnamefont {Ricci-Tersenghi}}, \bibinfo {author} {\bibfnamefont {M.}~\bibnamefont {Weigt}}, \ and\ \bibinfo {author} {\bibfnamefont {R.}~\bibnamefont {Zecchina}},\ }\href {\doibase 10.1103/PhysRevE.63.026702} {\bibfield  {journal} {\bibinfo  {journal} {Phys. Rev. E}\ }\textbf {\bibinfo {volume} {63}},\ \bibinfo {pages} {026702} (\bibinfo {year} {2001})}\BibitemShut {NoStop}%
\bibitem [{\citenamefont {Guidetti}\ and\ \citenamefont {Young}(2011)}]{guidetti2011complexity}%
  \BibitemOpen
  \bibfield  {author} {\bibinfo {author} {\bibfnamefont {M.}~\bibnamefont {Guidetti}}\ and\ \bibinfo {author} {\bibfnamefont {A.~P.}\ \bibnamefont {Young}},\ }\href {\doibase 10.1103/PhysRevE.84.011102} {\bibfield  {journal} {\bibinfo  {journal} {Phys. Rev. E}\ }\textbf {\bibinfo {volume} {84}},\ \bibinfo {pages} {011102} (\bibinfo {year} {2011})}\BibitemShut {NoStop}%
\bibitem [{\citenamefont {Garcia-Saez}\ and\ \citenamefont {Latorre}(2011)}]{garcia-saez_exact_2011}%
  \BibitemOpen
  \bibfield  {author} {\bibinfo {author} {\bibfnamefont {A.}~\bibnamefont {Garcia-Saez}}\ and\ \bibinfo {author} {\bibfnamefont {J.~I.}\ \bibnamefont {Latorre}},\ }\href@noop {} {\enquote {\bibinfo {title} {An exact tensor network for the {3SAT} problem},}\ } (\bibinfo {year} {2011}),\ \Eprint {http://arxiv.org/abs/1105.3201} {arXiv:1105.3201 [quant-ph]} \BibitemShut {NoStop}%
\bibitem [{\citenamefont {Biamonte}\ \emph {et~al.}(2015)\citenamefont {Biamonte}, \citenamefont {Morton},\ and\ \citenamefont {Turner}}]{biamonte2015tensor}%
  \BibitemOpen
  \bibfield  {author} {\bibinfo {author} {\bibfnamefont {J.~D.}\ \bibnamefont {Biamonte}}, \bibinfo {author} {\bibfnamefont {J.}~\bibnamefont {Morton}}, \ and\ \bibinfo {author} {\bibfnamefont {J.}~\bibnamefont {Turner}},\ }\href {\doibase 10.1007/s10955-015-1276-z} {\bibfield  {journal} {\bibinfo  {journal} {J. Stat. Phys.}\ }\textbf {\bibinfo {volume} {160}},\ \bibinfo {pages} {1389} (\bibinfo {year} {2015})}\BibitemShut {NoStop}%
\bibitem [{\citenamefont {Kourtis}\ \emph {et~al.}(2019)\citenamefont {Kourtis}, \citenamefont {Chamon}, \citenamefont {Mucciolo},\ and\ \citenamefont {Ruckenstein}}]{kourtis_fast_2019}%
  \BibitemOpen
  \bibfield  {author} {\bibinfo {author} {\bibfnamefont {S.}~\bibnamefont {Kourtis}}, \bibinfo {author} {\bibfnamefont {C.}~\bibnamefont {Chamon}}, \bibinfo {author} {\bibfnamefont {E.~R.}\ \bibnamefont {Mucciolo}}, \ and\ \bibinfo {author} {\bibfnamefont {A.~E.}\ \bibnamefont {Ruckenstein}},\ }\href {\doibase 10.21468/SciPostPhys.7.5.060} {\bibfield  {journal} {\bibinfo  {journal} {SciPost Phys.}\ }\textbf {\bibinfo {volume} {7}},\ \bibinfo {pages} {060} (\bibinfo {year} {2019})}\BibitemShut {NoStop}%
\bibitem [{\citenamefont {Meichanetzidis}\ and\ \citenamefont {Kourtis}(2019)}]{Meichanetzidis_2019}%
  \BibitemOpen
  \bibfield  {author} {\bibinfo {author} {\bibfnamefont {K.}~\bibnamefont {Meichanetzidis}}\ and\ \bibinfo {author} {\bibfnamefont {S.}~\bibnamefont {Kourtis}},\ }\href {\doibase 10.1103/PhysRevE.100.033303} {\bibfield  {journal} {\bibinfo  {journal} {Phys. Rev. E}\ }\textbf {\bibinfo {volume} {100}},\ \bibinfo {pages} {033303} (\bibinfo {year} {2019})}\BibitemShut {NoStop}%
\bibitem [{\citenamefont {de~Beaudrap}\ \emph {et~al.}(2021)\citenamefont {de~Beaudrap}, \citenamefont {Kissinger},\ and\ \citenamefont {Meichanetzidis}}]{de_beaudrap_tensor_2021}%
  \BibitemOpen
  \bibfield  {author} {\bibinfo {author} {\bibfnamefont {N.}~\bibnamefont {de~Beaudrap}}, \bibinfo {author} {\bibfnamefont {A.}~\bibnamefont {Kissinger}}, \ and\ \bibinfo {author} {\bibfnamefont {K.}~\bibnamefont {Meichanetzidis}},\ }\href {\doibase 10.4204/eptcs.340.3} {\bibfield  {journal} {\bibinfo  {journal} {EPTCS}\ }\textbf {\bibinfo {volume} {340}},\ \bibinfo {pages} {46–59} (\bibinfo {year} {2021})}\BibitemShut {NoStop}%
\bibitem [{\citenamefont {Schuch}\ \emph {et~al.}(2007)\citenamefont {Schuch}, \citenamefont {Wolf}, \citenamefont {Verstraete},\ and\ \citenamefont {Cirac}}]{schuch2007computational}%
  \BibitemOpen
  \bibfield  {author} {\bibinfo {author} {\bibfnamefont {N.}~\bibnamefont {Schuch}}, \bibinfo {author} {\bibfnamefont {M.~M.}\ \bibnamefont {Wolf}}, \bibinfo {author} {\bibfnamefont {F.}~\bibnamefont {Verstraete}}, \ and\ \bibinfo {author} {\bibfnamefont {J.~I.}\ \bibnamefont {Cirac}},\ }\href {\doibase 10.1103/PhysRevLett.98.140506} {\bibfield  {journal} {\bibinfo  {journal} {Phys. Rev. Lett.}\ }\textbf {\bibinfo {volume} {98}},\ \bibinfo {pages} {140506} (\bibinfo {year} {2007})}\BibitemShut {NoStop}%
\bibitem [{\citenamefont {Evenbly}\ and\ \citenamefont {Vidal}(2015)}]{evenbly2015tensor}%
  \BibitemOpen
  \bibfield  {author} {\bibinfo {author} {\bibfnamefont {G.}~\bibnamefont {Evenbly}}\ and\ \bibinfo {author} {\bibfnamefont {G.}~\bibnamefont {Vidal}},\ }\href {\doibase 10.1103/PhysRevLett.115.180405} {\bibfield  {journal} {\bibinfo  {journal} {Phys. Rev. Lett.}\ }\textbf {\bibinfo {volume} {115}},\ \bibinfo {pages} {180405} (\bibinfo {year} {2015})}\BibitemShut {NoStop}%
\bibitem [{\citenamefont {Evenbly}(2017)}]{evenbly2017algorithms}%
  \BibitemOpen
  \bibfield  {author} {\bibinfo {author} {\bibfnamefont {G.}~\bibnamefont {Evenbly}},\ }\href {\doibase 10.1103/PhysRevB.95.045117} {\bibfield  {journal} {\bibinfo  {journal} {Phys. Rev. B}\ }\textbf {\bibinfo {volume} {95}},\ \bibinfo {pages} {045117} (\bibinfo {year} {2017})}\BibitemShut {NoStop}%
\bibitem [{\citenamefont {Gray}\ and\ \citenamefont {Chan}(2024)}]{gray_hyper-optimized_2022}%
  \BibitemOpen
  \bibfield  {author} {\bibinfo {author} {\bibfnamefont {J.}~\bibnamefont {Gray}}\ and\ \bibinfo {author} {\bibfnamefont {G.~K.-L.}\ \bibnamefont {Chan}},\ }\href {\doibase 10.1103/PhysRevX.14.011009} {\bibfield  {journal} {\bibinfo  {journal} {Phys. Rev. X}\ }\textbf {\bibinfo {volume} {14}},\ \bibinfo {pages} {011009} (\bibinfo {year} {2024})}\BibitemShut {NoStop}%
\bibitem [{\citenamefont {Alkabetz}\ and\ \citenamefont {Arad}(2021)}]{alkabetz2021tensor}%
  \BibitemOpen
  \bibfield  {author} {\bibinfo {author} {\bibfnamefont {R.}~\bibnamefont {Alkabetz}}\ and\ \bibinfo {author} {\bibfnamefont {I.}~\bibnamefont {Arad}},\ }\href {\doibase 10.1103/PhysRevResearch.3.023073} {\bibfield  {journal} {\bibinfo  {journal} {Phys. Rev. Res.}\ }\textbf {\bibinfo {volume} {3}},\ \bibinfo {pages} {023073} (\bibinfo {year} {2021})}\BibitemShut {NoStop}%
\bibitem [{\citenamefont {Pancotti}\ and\ \citenamefont {Gray}(2023)}]{pancotti2023one}%
  \BibitemOpen
  \bibfield  {author} {\bibinfo {author} {\bibfnamefont {N.}~\bibnamefont {Pancotti}}\ and\ \bibinfo {author} {\bibfnamefont {J.}~\bibnamefont {Gray}},\ }\href@noop {} {\enquote {\bibinfo {title} {One-step replica symmetry breaking in the language of tensor networks},}\ } (\bibinfo {year} {2023}),\ \Eprint {http://arxiv.org/abs/2306.15004} {arXiv:2306.15004 [quant-ph]} \BibitemShut {NoStop}%
\bibitem [{\citenamefont {Garey}\ and\ \citenamefont {Johnson}(1979)}]{garey1979computers}%
  \BibitemOpen
  \bibfield  {author} {\bibinfo {author} {\bibfnamefont {M.~R.}\ \bibnamefont {Garey}}\ and\ \bibinfo {author} {\bibfnamefont {D.~S.}\ \bibnamefont {Johnson}},\ }\href@noop {} {\emph {\bibinfo {title} {Computers and Intractability: A Guide to the Theory of NP-Completeness}}}\ (\bibinfo  {publisher} {W. H. Freeman},\ \bibinfo {address} {San Francisco, CA},\ \bibinfo {year} {1979})\BibitemShut {NoStop}%
\bibitem [{\citenamefont {Braunstein}\ \emph {et~al.}(2002)\citenamefont {Braunstein}, \citenamefont {Leone}, \citenamefont {Ricci-Tersenghi},\ and\ \citenamefont {Zecchina}}]{braunstein_complexity_2002}%
  \BibitemOpen
  \bibfield  {author} {\bibinfo {author} {\bibfnamefont {A.}~\bibnamefont {Braunstein}}, \bibinfo {author} {\bibfnamefont {M.}~\bibnamefont {Leone}}, \bibinfo {author} {\bibfnamefont {F.}~\bibnamefont {Ricci-Tersenghi}}, \ and\ \bibinfo {author} {\bibfnamefont {R.}~\bibnamefont {Zecchina}},\ }\href {\doibase 10.1088/0305-4470/35/35/301} {\bibfield  {journal} {\bibinfo  {journal} {J. Phys. A: Math. Gen.}\ }\textbf {\bibinfo {volume} {35}},\ \bibinfo {pages} {7559} (\bibinfo {year} {2002})}\BibitemShut {NoStop}%
\bibitem [{\citenamefont {Denny}\ \emph {et~al.}(2011)\citenamefont {Denny}, \citenamefont {Biamonte}, \citenamefont {Jaksch},\ and\ \citenamefont {Clark}}]{denny_algebraically_2012}%
  \BibitemOpen
  \bibfield  {author} {\bibinfo {author} {\bibfnamefont {S.~J.}\ \bibnamefont {Denny}}, \bibinfo {author} {\bibfnamefont {J.~D.}\ \bibnamefont {Biamonte}}, \bibinfo {author} {\bibfnamefont {D.}~\bibnamefont {Jaksch}}, \ and\ \bibinfo {author} {\bibfnamefont {S.~R.}\ \bibnamefont {Clark}},\ }\href {\doibase 10.1088/1751-8113/45/1/015309} {\bibfield  {journal} {\bibinfo  {journal} {J. Phys. A: Math. Theor.}\ }\textbf {\bibinfo {volume} {45}},\ \bibinfo {pages} {015309} (\bibinfo {year} {2011})}\BibitemShut {NoStop}%
\bibitem [{\citenamefont {Seitz}\ \emph {et~al.}(2023)\citenamefont {Seitz}, \citenamefont {Medina}, \citenamefont {Cruz}, \citenamefont {Huang},\ and\ \citenamefont {Mendl}}]{seitz_simulating_2023}%
  \BibitemOpen
  \bibfield  {author} {\bibinfo {author} {\bibfnamefont {P.}~\bibnamefont {Seitz}}, \bibinfo {author} {\bibfnamefont {I.}~\bibnamefont {Medina}}, \bibinfo {author} {\bibfnamefont {E.}~\bibnamefont {Cruz}}, \bibinfo {author} {\bibfnamefont {Q.}~\bibnamefont {Huang}}, \ and\ \bibinfo {author} {\bibfnamefont {C.~B.}\ \bibnamefont {Mendl}},\ }\href {\doibase 10.22331/q-2023-03-30-964} {\bibfield  {journal} {\bibinfo  {journal} {{Quantum}}\ }\textbf {\bibinfo {volume} {7}},\ \bibinfo {pages} {964} (\bibinfo {year} {2023})}\BibitemShut {NoStop}%
\bibitem [{\citenamefont {Wang}\ \emph {et~al.}(2023)\citenamefont {Wang}, \citenamefont {Pan}, \citenamefont {Xu}, \citenamefont {Yang}, \citenamefont {Li},\ and\ \citenamefont {Cichocki}}]{wang_tensor_2023}%
  \BibitemOpen
  \bibfield  {author} {\bibinfo {author} {\bibfnamefont {M.}~\bibnamefont {Wang}}, \bibinfo {author} {\bibfnamefont {Y.}~\bibnamefont {Pan}}, \bibinfo {author} {\bibfnamefont {Z.}~\bibnamefont {Xu}}, \bibinfo {author} {\bibfnamefont {X.}~\bibnamefont {Yang}}, \bibinfo {author} {\bibfnamefont {G.}~\bibnamefont {Li}}, \ and\ \bibinfo {author} {\bibfnamefont {A.}~\bibnamefont {Cichocki}},\ }\href {\doibase 10.48550/arXiv.2302.09019} {\bibfield  {journal} {\bibinfo  {journal} {arXiv preprint arXiv:2302.09019}\ } (\bibinfo {year} {2023}),\ 10.48550/arXiv.2302.09019}\BibitemShut {NoStop}%
\bibitem [{\citenamefont {Gray}\ and\ \citenamefont {Kourtis}(2021)}]{gray_hyper-optimized_2021}%
  \BibitemOpen
  \bibfield  {author} {\bibinfo {author} {\bibfnamefont {J.}~\bibnamefont {Gray}}\ and\ \bibinfo {author} {\bibfnamefont {S.}~\bibnamefont {Kourtis}},\ }\href {\doibase 10.22331/q-2021-03-15-410} {\bibfield  {journal} {\bibinfo  {journal} {{Quantum}}\ }\textbf {\bibinfo {volume} {5}},\ \bibinfo {pages} {410} (\bibinfo {year} {2021})}\BibitemShut {NoStop}%
\bibitem [{\citenamefont {Viger}\ and\ \citenamefont {Latapy}(2015)}]{viger_efficient_2016}%
  \BibitemOpen
  \bibfield  {author} {\bibinfo {author} {\bibfnamefont {F.}~\bibnamefont {Viger}}\ and\ \bibinfo {author} {\bibfnamefont {M.}~\bibnamefont {Latapy}},\ }\href {\doibase 10.1093/comnet/cnv013} {\bibfield  {journal} {\bibinfo  {journal} {Journal of Complex Networks}\ }\textbf {\bibinfo {volume} {4}},\ \bibinfo {pages} {15} (\bibinfo {year} {2015})}\BibitemShut {NoStop}%
\bibitem [{\citenamefont {Gray}(2018)}]{gray2018quimb}%
  \BibitemOpen
  \bibfield  {author} {\bibinfo {author} {\bibfnamefont {J.}~\bibnamefont {Gray}},\ }\href {\doibase 10.21105/joss.00819} {\bibfield  {journal} {\bibinfo  {journal} {Journal of Open Source Software}\ }\textbf {\bibinfo {volume} {3}},\ \bibinfo {pages} {819} (\bibinfo {year} {2018})}\BibitemShut {NoStop}%
\bibitem [{\citenamefont {Csardi}\ and\ \citenamefont {Nepusz}(2006)}]{csardi_igraph_nodate}%
  \BibitemOpen
  \bibfield  {author} {\bibinfo {author} {\bibfnamefont {G.}~\bibnamefont {Csardi}}\ and\ \bibinfo {author} {\bibfnamefont {T.}~\bibnamefont {Nepusz}},\ }\href {https://igraph.org} {\bibfield  {journal} {\bibinfo  {journal} {InterJournal}\ }\textbf {\bibinfo {volume} {Complex Systems}},\ \bibinfo {pages} {1695} (\bibinfo {year} {2006})}\BibitemShut {NoStop}%
\bibitem [{\citenamefont {Girvan}\ and\ \citenamefont {Newman}(2002)}]{girvan_community_2002}%
  \BibitemOpen
  \bibfield  {author} {\bibinfo {author} {\bibfnamefont {M.}~\bibnamefont {Girvan}}\ and\ \bibinfo {author} {\bibfnamefont {M.~E.~J.}\ \bibnamefont {Newman}},\ }\href {\doibase 10.1073/pnas.122653799} {\bibfield  {journal} {\bibinfo  {journal} {Proc. Natl. Acad. Sci.}\ }\textbf {\bibinfo {volume} {99}},\ \bibinfo {pages} {7821} (\bibinfo {year} {2002})}\BibitemShut {NoStop}%
\bibitem [{\citenamefont {Schlag}\ \emph {et~al.}()\citenamefont {Schlag}, \citenamefont {Henne}, \citenamefont {Heuer}, \citenamefont {Meyerhenke}, \citenamefont {Sanders},\ and\ \citenamefont {Schulz}}]{kahypar2016}%
  \BibitemOpen
  \bibfield  {author} {\bibinfo {author} {\bibfnamefont {S.}~\bibnamefont {Schlag}}, \bibinfo {author} {\bibfnamefont {V.}~\bibnamefont {Henne}}, \bibinfo {author} {\bibfnamefont {T.}~\bibnamefont {Heuer}}, \bibinfo {author} {\bibfnamefont {H.}~\bibnamefont {Meyerhenke}}, \bibinfo {author} {\bibfnamefont {P.}~\bibnamefont {Sanders}}, \ and\ \bibinfo {author} {\bibfnamefont {C.}~\bibnamefont {Schulz}},\ }\enquote {\bibinfo {title} {\emph{k}-way hypergraph partitioning via \emph{n}-level recursive bisection},}\ in\ \href {\doibase 10.1137/1.9781611974317.5} {\emph {\bibinfo {booktitle} {2016 Proceedings of the Meeting on Algorithm Engineering and Experiments (ALENEX)}}},\ pp.\ \bibinfo {pages} {53--67},\ \Eprint {http://arxiv.org/abs/https://epubs.siam.org/doi/pdf/10.1137/1.9781611974317.5} {https://epubs.siam.org/doi/pdf/10.1137/1.9781611974317.5} \BibitemShut {NoStop}%
\bibitem [{\citenamefont {Akhremtsev}\ \emph {et~al.}()\citenamefont {Akhremtsev}, \citenamefont {Heuer}, \citenamefont {Sanders},\ and\ \citenamefont {Schlag}}]{kahypar2017}%
  \BibitemOpen
  \bibfield  {author} {\bibinfo {author} {\bibfnamefont {Y.}~\bibnamefont {Akhremtsev}}, \bibinfo {author} {\bibfnamefont {T.}~\bibnamefont {Heuer}}, \bibinfo {author} {\bibfnamefont {P.}~\bibnamefont {Sanders}}, \ and\ \bibinfo {author} {\bibfnamefont {S.}~\bibnamefont {Schlag}},\ }\enquote {\bibinfo {title} {Engineering a direct \emph{k}-way hypergraph partitioning algorithm},}\ in\ \href {\doibase 10.1137/1.9781611974768.3} {\emph {\bibinfo {booktitle} {2017 Proceedings of the Meeting on Algorithm Engineering and Experiments (ALENEX)}}},\ pp.\ \bibinfo {pages} {28--42},\ \Eprint {http://arxiv.org/abs/https://epubs.siam.org/doi/pdf/10.1137/1.9781611974768.3} {https://epubs.siam.org/doi/pdf/10.1137/1.9781611974768.3} \BibitemShut {NoStop}%
\bibitem [{\citenamefont {Zhu}\ and\ \citenamefont {Katzgraber}(2019)}]{zhu2019tensor}%
  \BibitemOpen
  \bibfield  {author} {\bibinfo {author} {\bibfnamefont {Z.}~\bibnamefont {Zhu}}\ and\ \bibinfo {author} {\bibfnamefont {H.~G.}\ \bibnamefont {Katzgraber}},\ }\href {\doibase 10.48550/arXiv.1903.07721} {\bibfield  {journal} {\bibinfo  {journal} {arXiv preprint arXiv:1903.07721}\ } (\bibinfo {year} {2019}),\ 10.48550/arXiv.1903.07721}\BibitemShut {NoStop}%
\bibitem [{\citenamefont {Fattal}\ \emph {et~al.}(2004)\citenamefont {Fattal}, \citenamefont {Cubitt}, \citenamefont {Yamamoto}, \citenamefont {Bravyi},\ and\ \citenamefont {Chuang}}]{fattal2004entanglement}%
  \BibitemOpen
  \bibfield  {author} {\bibinfo {author} {\bibfnamefont {D.}~\bibnamefont {Fattal}}, \bibinfo {author} {\bibfnamefont {T.~S.}\ \bibnamefont {Cubitt}}, \bibinfo {author} {\bibfnamefont {Y.}~\bibnamefont {Yamamoto}}, \bibinfo {author} {\bibfnamefont {S.}~\bibnamefont {Bravyi}}, \ and\ \bibinfo {author} {\bibfnamefont {I.~L.}\ \bibnamefont {Chuang}},\ }\href {\doibase 10.48550/arXiv.quant-ph/0406168} {\bibfield  {journal} {\bibinfo  {journal} {arXiv preprint quant-ph/0406168}\ } (\bibinfo {year} {2004}),\ 10.48550/arXiv.quant-ph/0406168}\BibitemShut {NoStop}%
\bibitem [{\citenamefont {Hamma}\ \emph {et~al.}(2005)\citenamefont {Hamma}, \citenamefont {Ionicioiu},\ and\ \citenamefont {Zanardi}}]{Hamma2005entropy}%
  \BibitemOpen
  \bibfield  {author} {\bibinfo {author} {\bibfnamefont {A.}~\bibnamefont {Hamma}}, \bibinfo {author} {\bibfnamefont {R.}~\bibnamefont {Ionicioiu}}, \ and\ \bibinfo {author} {\bibfnamefont {P.}~\bibnamefont {Zanardi}},\ }\href {\doibase 10.1103/PhysRevA.71.022315} {\bibfield  {journal} {\bibinfo  {journal} {Phys. Rev. A}\ }\textbf {\bibinfo {volume} {71}},\ \bibinfo {pages} {022315} (\bibinfo {year} {2005})}\BibitemShut {NoStop}%
\bibitem [{\citenamefont {Zhou}\ \emph {et~al.}(2020)\citenamefont {Zhou}, \citenamefont {Yang}, \citenamefont {Hamma},\ and\ \citenamefont {Chamon}}]{zhou2020Clifford}%
  \BibitemOpen
  \bibfield  {author} {\bibinfo {author} {\bibfnamefont {S.}~\bibnamefont {Zhou}}, \bibinfo {author} {\bibfnamefont {Z.-C.}\ \bibnamefont {Yang}}, \bibinfo {author} {\bibfnamefont {A.}~\bibnamefont {Hamma}}, \ and\ \bibinfo {author} {\bibfnamefont {C.}~\bibnamefont {Chamon}},\ }\href {\doibase 10.21468/SciPostPhys.9.6.087} {\bibfield  {journal} {\bibinfo  {journal} {SciPost Phys.}\ }\textbf {\bibinfo {volume} {9}},\ \bibinfo {pages} {087} (\bibinfo {year} {2020})}\BibitemShut {NoStop}%
\end{thebibliography}%

\end{document}